\documentclass[pra,eqsecnum,twocolumn,amsmath,amssymb]{revtex4-2}

\usepackage{graphicx}% Include figure files
\usepackage{bm}% bold math
\usepackage[usenames,dvipsnames]{color}
\definecolor{altncolor}{rgb}{0,0,0.8}
\usepackage[colorlinks=true, linkcolor=green, anchorcolor=altncolor,
citecolor=altncolor, filecolor=altncolor, menucolor=altncolor,
urlcolor=altncolor]{hyperref}

\begin{document}

\title{Statistical equilibrium principles in 2D fluid flow: from geophysical fluids to the solar tachocline}
% Force line breaks with \\

\author{Peter B. Weichman$^1$ and J. B. Marston$^2$}

\affiliation{$^1$FAST Labs, BAE Systems, Technology Solutions, 600 District
Avenue, Burlington, MA 01803 USA\\
$^2$Brown Theoretical Physics Center and Department of Physics, Brown University, Providence, RI 02912-S USA}

% \date{\today}

\begin{abstract}

An overview is presented of several diverse branches of work in the area of effectively 2D fluid equilibria which have in common that they are constrained by an infinite number of conservation laws. Broad concepts, and the enormous variety of physical phenomena that can be explored, are highlighted. These span, roughly in order of increasing complexity, Euler flow, nonlinear Rossby waves, 3D axisymmetric flow, shallow water dynamics, and 2D magnetohydrodynamics. The classical field theories describing these systems bear some resemblance to perhaps more familiar fluctuating membrane and continuous spin models, but the fluid physics drives these models into unconventional regimes exhibiting large scale jet and eddy structures. From a dynamical point of view these structures are the end result of various conserved variable forward and inverse cascades. The resulting balance between large scale structure and small scale fluctuations is controlled by the competition between energy and entropy in the system free energy, in turn highly tunable through setting the values of the conserved integrals. Although the statistical mechanical description of such systems is fully self-consistent, with remarkable mathematical structure and diversity of solutions, great care must be taken because the underlying assumptions, especially ergodicity, can be violated or at minimum lead to exceedingly long equilibration times. Generalization of the theory to include weak driving and dissipation (e.g., non-equilibrium statistical mechanics and associated linear response formalism) could provide additional insights, but has yet to be properly explored.

\end{abstract}

%\pacs{47.10.1g, 05.70.Ln, 05.90.1m, 52.30.2q}
%\pacs{52.35.-g, 52.35.Kt, 52.35.Mw, 52.35.Ra, 47.27.De}
                                   % Classification Scheme (http://www.aip.org/pacs/)
\maketitle

\section{Introduction}
\label{sec:intro}

Remarkable progress has been made over the past 30 years or so applying rigorous statistical equilibrium principles to classical fluid systems with increasing degrees of complexity \cite{BV2012}. The essential idea is that a freely decaying, strongly turbulent initial condition at late time is often observed to relax into a macroscopically smooth steady state (illustrated in Fig.\ \ref{fig:turbmix}). These ideas are especially interesting in two dimensions where \emph{inverse} cascades can generate nontrivial macroscopic features, such as system-spanning eddies or jets, from purely small scale, but significantly nonlinear fluctuations. Moreover, additional strong constraints, that forbid 2D eddies from ``turning over'' and effectively self-canceling, lead to an infinite number of additional conserved integrals of the motion, known as Casimirs. Acting together, all of these lead to a similarly infinite number of possible late-time flow geometries. These are exemplified, e.g., by Jupiter's Great Red Spot, gas giant latitudinal band structure, polar vortices \cite{SaturnPV1988,SaturnPV2018,JupiterPV2020}, and other planetary flows.

Some, but by no means all, of these near-steady state long-lived, structures might be considered as weakly driven, balanced by weak dissipation. It then becomes interesting to seek quantitative and qualitative insights using models in an idealized zero driving, zero dissipation limit. For realistic comparisons, these models may additionally require nontrivial multilayer vertical structure. Here we consider only the simplest models with the fewest number of degrees of freedom exhibiting various fundamental physical phenomena. These ideas have a very long history \cite{O1949,K1975,MJ1974,LP1977,LB1967} but notable progress was made in the early 1990s \cite{M1990,RS1991,MWC1992,MR1994}, motivated in part by then recent numerical simulation results \cite{Marcus1988,Marcus1990}, accounting for the full set of Casimirs in place of, e.g., point vortex and energy--enstrophy (Gaussian model) approximations.

From a mathematical point of view, the statistical mechanics of fluid motions ultimately reduces to study of certain classical field theories, especially involving the velocity field and fields derived from it. Some of these theories bear resemblance to those describing more conventional systems, such as Ising spin models and fluctuating elastic membranes. However, the most interesting fluid behaviors, especially those pertaining to large scale flows, tend to correspond to unusual limits of these models that have not been previously explored. Table \ref{tab:fields} provides a summary of these field theories that will be explained in much greater detail in later sections.

Predictions for the late time equilibrium state, assuming that it is reached, are based only on certain macroscopic features of the initial condition, namely the values of the conserved integrals, including total energy, linear or angular momentum, and the Casimirs. Although insensitive to the details of the turbulent decay that gives rise to these states, such predictions, beyond their intrinsic interest, could provide useful consistency checks on results from late time direct numerical simulations. Conversely, lack of consistency, if indeed robustly borne out by the numerics, could point to existence interesting equilibration barriers and metastable behaviors. There is already significant evidence that such barriers are much more common in such highly constrained 2D flows than in, e.g., conventional particle systems, through a variety of mechanisms \cite{BNZ1991,BHW2011,CC1996,CC1996b,QM2014,DQM2015,D2020}.

The remainder of this paper is summarized as follows. We begin by presenting a fairly detailed derivation of the statistical equilibrium theory for the simplest possible model, the 2D Euler equation, which is fully described by the scalar vorticity. The ingredients of this theory follow a logical chain that is repeated, or extended as necessary, for the more complicated systems. In Sec.\ \ref{sec:2deuler} the equations of motion are introduced and their reduction to the vorticity field dynamics demonstrated. The usual energy and momentum conservation laws are exhibited, followed by the Casimir constraints.

General equilibrium concepts are introduced in Sec.\ \ref{sec:stateq} in terms of invariant (steady state) measures over the phase space of all vorticity configurations. Identifying such measures relies on the Liouville theorem, which establishes a type of phase space incompressibility condition. Once proven, the allowed measures are constructed from the fluid conserved integrals themselves, and the exact choice corresponds to what is known as a statistical ensemble. The thermodynamic entropy, free energy, etc., follow from the logarithm of the measure phase space integral (partition function) in the usual way. The grand canonical ensemble for the Euler equation is introduced as providing the most convenient mathematical framework.

The general statistical formalism is applied to the 2D Euler in Sec.\ \ref{sec:tdeulerexact}. Perhaps surprisingly, given the infinite number of constraints, the system free energy may actually be derived \emph{exactly} as an explicit variational equation---the long range Coulomb-like vortex interactions enable an exact mean field-type approximation \cite{M1990,RS1991,MWC1992}. The minima describe the various possible equilibrium states, whose large scale flow pattern varies with the specified conserved integral values. Critically, the Casimir constraints permit both positive and negative temperature equilibria, with the latter encouraging compact eddy structures reminiscent of Jupiter's Red Spot.  There is again a very interesting competition between energy and entropy that controls the amplitude and size of such structures. Simple two-level system models are introduced that allow convenient exploration of these phenomena.

A brief discussion of some of the limitations of the statistical equilibrium hypothesis is presented in Sec.\ \ref{sec:eqhyplimits}. Vortex mixing dynamics in 2D is clearly far more constrained than particle dynamics underlying conventional systems (though microscale viscosity, neglected here, in a sense bridges the two regimes). It should therefore not be too surprising that significant barriers to equilibration can occur \cite{BV2012}. Some of these barriers can actually be understood as local rather than global minima of the free energy functional. Examples include separated compact eddies that orbit each other, failing to merge (as would be entropically favored) above a critical separation \cite{CC1996}. Detailed numerical simulations show evidence for different levels of equilibration in different spatial regions, depending on the strength of local mixing dynamics \cite{CC1996b}. Others are somewhat more mysterious: equilibration on the surface of a sphere (rather than in a flat bounded domain) is found to fail much more catastrophically, with a macroscopically fluctuating chaotic vorticity field surviving for all achievable computation times \cite{DQM2015}.

In Sec.\ \ref{sec:statgen1d} we discuss the most straightforward generalization of the Euler results to a more general class of single scalar field systems whose canonical structure automatically ensure an infinite set of Casimirs. Under reasonable conditions, mean field approximation is again exact, and the free energy functional emerges from a Legendre transformation of the energy. An important example is the quasigeostrophic (QG) equation, a scalar field approximation to the shallow water equations \cite{BS2002,W2006}. This system also has an additional approximate adiabatic invariant \cite{BNZ1991} that is completely separate from the standard conservation laws, provides another possible equilibration barrier example.

In Sec.\ \ref{sec:3daxisym} we consider 3D axisymmetric flow in which azimuthal symmetry is imposed on flows confined to cylinder (Taylor--Couette geometry). The equations of motion now reduce to a coupled pair of scalar equations describing coupled toroidal and poloidal flow, with only the former experiencing the Casimir constraints \cite{HMRW1985}. However it is the poloidal flow, within each range--height slice, that is most directly analogous to the Euler equation vorticity. The fact that it is now only indirectly influenced by the Casimirs drastically changes the character of the equilibrium state \cite{LDC2006,NTCCD2010,NMCD2010,TDB2014,W2019}. The poloidal vorticity exhibits no large scale structure, though the velocity field does maintain strong microscale fluctuations. The toroidal velocity field exhibits relatively simple radial band-like structure controlled by the Casimirs \cite{W2019}.

In Sec.\ \ref{sec:swwaveeddy} we consider the full shallow water equations, which may be reduced to three coupled scalar equations, with again only one of them, the potential vorticity equation, possessing Casimir constraints. The statistical fluctuations of both compressional part of the velocity and the surface height remain very strong in equilibrium, and these drive similarly strong fluctuations in the vortex interactions \cite{W2017}, playing the role of an unbounded heat sink that precludes the existence of negative temperature eddy-like states \cite{W2017,RVB2016}. This raises very interesting questions, which cannot be answered by an equilibrium theory alone, regarding the rate at which wave--eddy interactions dissipate such structures if they are created in the initial state, and how they might be maintained (as seen in planetary atmospheres and in experiments) outside of equilibrium. Most optimistically, there may be mechanisms by which additional weak dissipation processes, such as wave breaking, can act to differentially suppress the waves, maintaining the eddies as formally metastable near-equilibria. We exhibit a possible variational formalism, a fairly straightforward generalization of that describing Euler and QG equilibria, that might be used to approximately describe these \cite{WP2001,CS2002}. This system also has a separate adiabatic invariant \cite{BHW2011}.

In Sec.\ \ref{sec:2Dmhd} we consider magnetohydrodynamic flow of perfectly conducting fluids, which couple mass and electrical current flow through the Maxwell equations. This model has been used to model the solar tachocline \cite{TDH2007,PMT2019} which marks the very thin 2D boundary between the rigidly rotating radiative interior and the differentially rotating exterior convective zone. The results here are significantly different than all previous examples because the Casimir constraints are tied to the magnetic vector potential instead of the vorticity \cite{HMRW1985,JT1997,LDC2005}. The model that emerges maps onto a pair of interacting elastic membranes in an external confining potential controlled by the Casimirs \cite{W2012}. The microscale fluctuations are purely Gaussian, and this allows a formally exact derivation of the free energy functional whose minima again determine the large scale structure of the magnetic and flow fields. In the solar context, the structure of these fields has implications for the transport of angular momentum between the two zones.

The paper is concluded in Sec.\ \ref{sec:conclude}. It is remarkable how much physical structure the equilibrium theories contain, and how different this structure is for each of the examples treated. There are a number of other well known systems with Casimir constraints \cite{HMRW1985} that can still be explored. Near-equilibrium generalizations are also of great interest.

\begin{table*}

\begin{tabular}{l|ll}
2D Euler [Secs.\ \ref{sec:2deuler}--\ref{sec:eqhyplimits}; vorticity $\omega$,
planar coordinate ${\bf r} = (x,y)$]
& &
${\cal K}[\omega] = \frac{1}{2} \int d{\bf r} \int d{\bf r}'
\omega({\bf r}) G({\bf r},{\bf r}') \omega({\bf r}')$ \\
\ \ \ \ and QG flow [Sec.\ \ref{sec:statgen1d}; potential vorticity $\omega$]
\\
Continuous spin, long range interacting Ising-type model, with spin
& &
$\ \ \ \ - \int d{\bf r} \left\{\mu_P \alpha({\bf r}) \omega({\bf r})
+ \mu[\omega({\bf r})] \right\}$
\\
\ \ \ \ weighting function $\mu(\omega)$, inhomogeneous magnetic field
$\mu_P \alpha({\bf r})$
& & \\
\\ \hline
\\
3D Axisymmetric flow [Sec.\ \ref{sec:3daxisym}; poloidal vorticity $q$, toroidal
& &
${\cal K}[q,s] = \pi \int d{\bm \rho} \int d{\bm \rho}'
q({\bf r}) G({\bm \rho},{\bm \rho}') q({\bm r}')$
\\
\ \ \ \ circulation $s$, radial-vertical ``cylinder slice'' coordinate
& & \\
\ \ \ \ ${\bm \rho} = (\rho_1,\rho_2) = (r^2/2, z)$]
& &
$\ \ \ \ + \int d{\bm \rho} \left\{\frac{\pi}{2\rho_1} s({\bm \rho})^2
- \mu[s({\bm \rho})] - \tilde \mu[s({\bm \rho})] q({\bm \rho}) \right\}$
\\
Continuous Ising-type spin $s$, weighting function $\mu(s)$,
& & \\
\ \ \ \ mediated by Gaussian ``charge'' field $q$,
coupling strength $\tilde \mu(s)$
& & \\
\\ \hline
\\
Shallow water equations [Sec.\ \ref{sec:swwaveeddy};
vorticity $\omega$, potential vorticity $\omega/h$,
& &
${\cal K}[\omega,q,h] = \frac{1}{2} \int d{\bf r} \int d{\bf r}'
\left[\begin{array}{c} \omega({\bf r}) \\ q({\bf r}) \end{array} \right]^T
{\cal G}_h({\bf r},{\bf r}')
\left[\begin{array}{c} \omega({\bf r}') \\ q({\bf r}') \end{array} \right]$
\\
\ \ \ \ compression field $q$, surface height $h$,
planar coordinate ${\bf r} = (x,y)$]
& &
\\
Continuous spin, long-range interacting Ising-type field $\omega$, with spin
& &
$\ \ \ \ + \int d{\bf r} \left\{\frac{1}{2} g h({\bf r})^2
- h({\bf r}) \mu[\omega({\bf r})/h({\bf r})] \right\}$
\\
\ \ \ \ weighting $\mu(\omega/h)$, tensor-coupled nonlinearly to Gaussian fields $q,h$
& & \\
\\ \hline
\\
2D magnetohydrodynamics [Sec.\ \ref{sec:2Dmhd}; stream function $\psi$,
& &
${\cal K}[A,\psi] = \int d{\bf r} \big\{\frac{1}{2} |\nabla A({\bf r})|^2
+ \frac{1}{2} |\nabla \psi({\bf r})|^2$
\\
\ \ \ \ magnetic vector potential $A$, planar section ${\bf r} = (x,y)$
& & \\
\ \ \ \ orthogonal to electric current density
$J = -\nabla^2 A$ along $\hat {\bf z}$]
& &
$\ \ \ \ -\ \tilde \mu'[A({\bf r})] \nabla A({\bf r}) \cdot \nabla \psi({\bf r})
+ \mu[A({\bf r})] \big\}$
\\
Model is equivalent to that of a pair of gradient-coupled elastic
& & \\
\ \ \ \ membranes, external confining and coupling potentials $\mu(A),\tilde \mu(A)$
& & \\
\end{tabular}

\caption{Summary of the (classical) field theoretic formulations of the statistical mechanics of the fluid systems discussed in this article. Statistical averages are governed by a phase space equilibrium probability density $\rho_\mathrm{eq} = Z_\mathrm{eq}^{-1} e^{-\beta {\cal K}[{\bm \Phi}]}$ in which $\beta = 1/T$ is an inverse temperature variable (quite distinct from the physical temperature) conjugate to the system entropy. The partition function $Z_\mathrm{eq}$ normalizes $\rho_\mathrm{eq}$ to a probability. The statistical functional ${\cal K}[{\bm \Phi}]$ is a certain conserved integral of the system dynamics, generalizing the energy (or Hamiltonian) familiar from conventional statistical mechanics. Its argument is a vector field ${\bm \Phi}({\bf r})$ whose components are the fluid degrees of freedom, and where ${\bf r}$ is a 2D physical spatial coordinate (sometimes scaled or transformed) ranging over a finite domain ${\cal D}$. The allowed configurations of ${\bm \Phi}$ define the thermodynamic phase space. The table summarizes the forms of ${\cal K}$ and ${\bm \Phi}$ for the systems treated in the referenced sections of the paper. The Coriolis parameter $f({\bf r})$, through which rotation enters, has been left out of these forms for simplicity, but will be restored in the later sections.}

\label{tab:fields}
\end{table*}

\section{Two-dimensional Euler equation}
\label{sec:2deuler}

It is useful to consider first the simplest system, the 2D Euler equation \cite{M1990,RS1991,MWC1992} defined by the equation of motion
\begin{equation}
\frac{D {\bf v}}{Dt} \equiv \partial_t {\bf v}
+ ({\bf v} \cdot \nabla) {\bf v} = -\nabla p
\label{2.1}
\end{equation}
in some 2D domain ${\cal D}$. The pressure $p$ is determined by the incompressibility constraint
\begin{equation}
\nabla \cdot {\bf v} = 0.
\label{2.2}
\end{equation}

\subsection{Vorticity and stream function}
\label{sec:vorticitystream}

\begin{figure}
\includegraphics[width=3.2in,viewport = 160 100 850 420,clip]{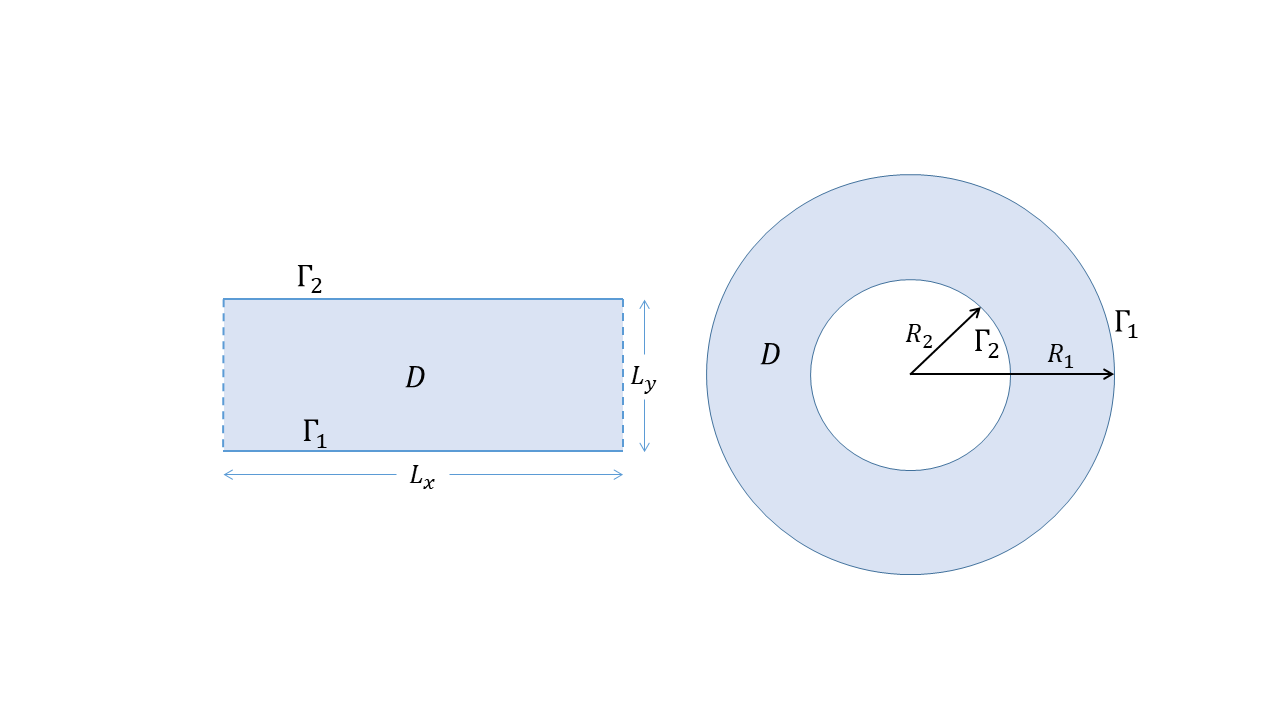}

\caption{Strip and annular (or disc if $R_2 = 0$) geometries for which, respectively, a conserved linear momentum (\ref{2.10}) or angular momentum (\ref{2.11}) exists. The strip has periodic boundary conditions along $x$ and Dirichlet boundary conditions on the lower and upper boundaries $\Gamma_{1,2}$. The annulus has Dirichlet boundary conditions on both boundaries. The latter lead to two independent circulation integrals (\ref{2.8}) for each domain (which are seen to actually have the same topology).}

\label{fig:momentumconserve}
\end{figure}

The constraint (\ref{2.1}) permits the stream function representation
\begin{equation}
{\bf v} = \nabla \times \psi \equiv (\partial_y, -\partial_x) \psi.
\label{2.3}
\end{equation}
By taking the curl of both sides of (\ref{2.1}), one obtains the vorticity equation
\begin{eqnarray}
\frac{D \omega}{Dt} &\equiv& \partial_t \omega + {\bf v} \cdot \nabla \omega = 0
\nonumber \\
\omega &=& \nabla \times {\bf v} \equiv \partial_x v_y - \partial_y v_x
\label{2.4}
\end{eqnarray}
which physically states that $\omega$ is freely advected by its own induced velocity field ${\bf v}$, constructed below. From (\ref{2.3}) follows the relation
\begin{equation}
\omega = -\nabla^2 \psi
\label{2.5}
\end{equation}
with formal solution
\begin{equation}
\psi({\bf r}) = \int_{\cal D} d{\bf r}' G({\bf r},{\bf r}') \omega({\bf r}'),
\label{2.6}
\end{equation}
in which the Laplace Green function is the solution to
\begin{equation}
-\nabla^2 G({\bf r},{\bf r}') = \delta({\bf r}-{\bf r}')
\label{2.7}
\end{equation}
together with the same boundary conditions, on both ${\bf r}$ and ${\bf r}'$, satisfied by $\psi$. Energy conservation requires free slip boundary conditions, equivalent to constant $\psi$ (Dirichlet boundary conditions). If there are multiple boundaries $\Gamma_n$, $n=1,2,\ldots,n_\partial$, e.g., an annular (see Fig.\ \ref{fig:momentumconserve}) or more general multi-holed domain, then $\psi = \psi_{0,n}$ may be assigned separate values on each boundary and are also constants of the motion. The circulation
\begin{equation}
\gamma_n = \int_{\Gamma_n} {\bf v} \cdot d{\bf l}
\label{2.8}
\end{equation}
about each boundary is also conserved. However, since the constants $\psi_{n,0}$ uniquely define $\psi$, it follows that the $\gamma_n$ are not independently conserved, but are (linearly) related to the former.

Equations (\ref{2.3}) and (\ref{2.5}) together uniquely determine ${\bf v}$ in terms of $\omega$, so that the first line of (\ref{2.4}) indeed represents a (scalar) closed evolution equation.

\subsection{Conservation laws}
\label{sec:conservelaws}

The conserved energy is just the kinetic energy
\begin{eqnarray}
E &=& \frac{1}{2} \int_{\cal D} d{\bf r} |{\bf v}({\bf r})|^2
= \frac{1}{2} \int_{\cal D} d{\bf r} |\nabla \psi({\bf r})|^2
\nonumber \\
&=& \frac{1}{2} \int_{\cal D} d{\bf r} \psi({\bf r}) \omega({\bf r})
\nonumber \\
&=& \frac{1}{2} \int_{\cal D} d{\bf r} \int_{\cal D} d{\bf r}'
\omega({\bf r}) G({\bf r},{\bf r}') \omega({\bf r}')
\label{2.9}
\end{eqnarray}
in which the boundary conditions ensure absence of boundary terms in the integration by parts used to obtain the second line, and (\ref{2.6}) has then been substituted to obtain the last line.

If the domain is translation invariant along some direction $\hat {\bf l}$ (infinite or periodic strip geometry, illustrated on the left in Fig.\ \ref{fig:momentumconserve}) then the corresponding component of the linear momentum
\begin{equation}
P_{\hat{\bf l}} = \int_{\cal D} d{\bf r} \hat {\bf l} \cdot {\bf v}
= \int_{\cal D} d{\bf r} \hat {\bf l} \times {\bf r} \omega({\bf r})
\label{2.10}
\end{equation}
is conserved. If the domain is rotation invariant (disc or annular geometry, illustrated on the right in Fig.\ \ref{fig:momentumconserve}) then the vertical component of the angular momentum
\begin{equation}
L_z = \int d{\bf r} {\bf r} \times {\bf v}
= \frac{1}{2} \int d{\bf r} r^2 \omega({\bf r})
\label{2.11}
\end{equation}
is conserved. Both of these can be written in the linear form
\begin{equation}
P = \int_{\cal D} \alpha({\bf r}) \omega({\bf r})
\label{2.12}
\end{equation}
with the choice $\alpha({\bf r}) = \hat {\bf l} \times {\bf r}$ or $\frac{1}{2} r^2$, depending on the domain. Note that on a true spherical domain, the full vector angular momentum ${\bf L}$ is conserved and (\ref{2.11}) is generalized appropriately.

The ``self-advection'' equation (\ref{2.4}) implies that any (1D) function of the vorticity
\begin{equation}
\Omega_F = \int_{\cal D} d{\bf r} F[\omega({\bf r})]
\label{2.13}
\end{equation}
is conserved. These may be conveniently summarized by conservation of the function
\begin{equation}
g(\sigma) = \int_{\cal D} d{\bf r} \delta[\sigma - \omega({\bf r})]
\label{2.14}
\end{equation}
for any value of $\sigma$, in terms of which
\begin{equation}
\Omega_F = \int d\sigma F(\sigma) g(\sigma).
\label{2.15}
\end{equation}
These are very often exhibited in terms of the powers $F(\sigma) = \sigma^n$, which are seen to generate the moments of $g(\sigma)$.

\begin{figure}

\includegraphics[width=3.2in,viewport = 40 140 910 380,clip]{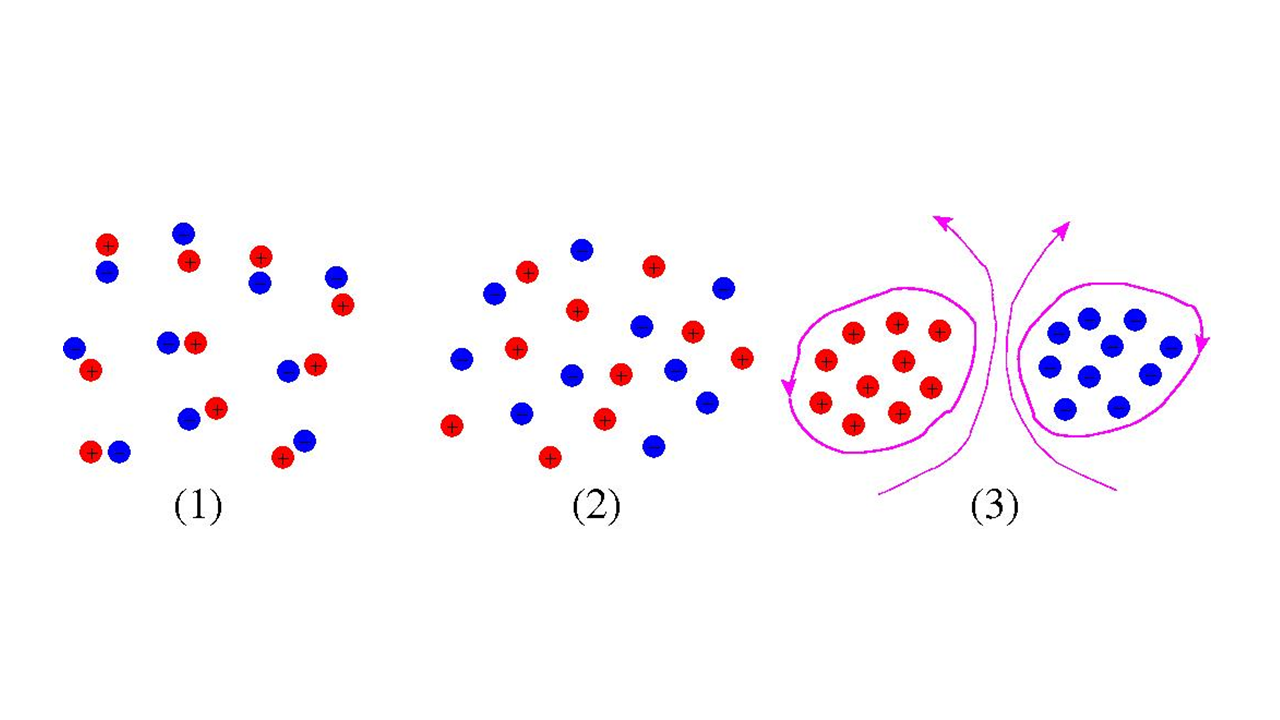}

\caption{Notional illustration of the equilibrium state of an overall neutral set of point-like vortices or charges as a function of internal energy \cite{O1949}. The low energy state on the left corresponds to a molecular dipole state with strongly bound charges. The middle state corresponds to a higher energy plasma-like state with unbounded charges but that continue to obey local charge neutrality. The state on the right exhibits large scale structure obtained by increasing the energy even further, forcing the charges to segregate into separate nonneutral regions. This negative temperature state is accessible in fluid dynamics because the charges are not conventional momentum and kinetic energy carrying particles. In the vortex field description, charges carry only potential energy of interaction.}

\label{fig:ptvortex}
\end{figure}

\section{Statistical equilibrium concepts}
\label{sec:stateq}

We now summarise the key statistical equilibrium concepts underlying the thermodynamic fluid treatment, especially the key role of microscale entropy. These concepts will serve to define the mathematical basis for computing thermodynamic functions and using them to characterize large scale steady state flows and other quantities of physical interest. A notional picture is illustrated in Fig.\ \ref{fig:ptvortex}, going back to the original ideas of Onsager \cite{O1949}. Conventional positive temperature bound ``molecule'' (left) and unbound plasma states (middle) exhibit no large scale vorticity or flow structure. In this picture, the physically interesting fluid equilibria correspond to much higher energy flows (right) in which the charges are forced to segregate, effectively like attracting like. We will see that such states indeed emerge as \emph{negative} temperature equilibria.

\begin{figure}

\includegraphics[width=3.2in,viewport = 260 80 690 460,clip]{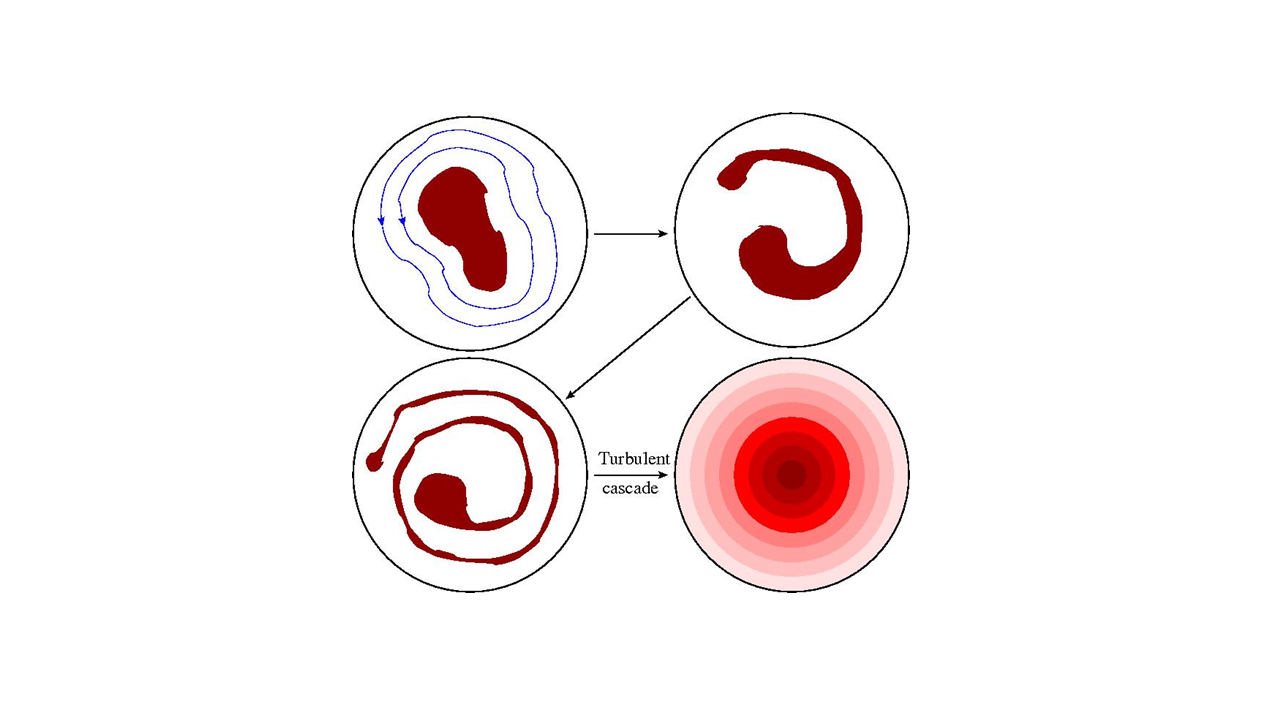}

\caption{Highly schematic illustration of the turbulent mixing process that begins here with a well defined though irregular region of finite, fixed vorticity $\omega = q_0$, surrounded by a vorticity free (potential flow) region, $\omega = 0$. Over time the vortex region stretches and folds to give rise as $t \to \infty$ to a fully mixed smoothly varying macroscale steady state. However, the macro-view obscures the continuing microscale dynamics (illustrated in Fig.\ \ref{fig:micromacro}) where restriction to values $\omega = 0,q$ is preserved, consistent with the Casimir constraints.}

\label{fig:turbmix}
\end{figure}

The standard underlying assumption, known as the ergodic hypothesis, is that very long time averages beginning from some given initial condition are equivalent to certain phase space averages over all field configurations consistent with the conservation laws. Figure \ref{fig:turbmix} schematically illustrates this idea for Euler flow, in which the turbulent mixing process eventually produces a smooth looking steady state with the original discrete vorticity levels hidden at the finest scales.

This section will detail the phase space averaging process under the ergodic hypothesis. Ergodicity is almost never provable from first principles and it can indeed be violated even in conventional particle systems. As discussed in Sec.\ \ref{sec:eqhyplimits}, and hinted at in other sections, violations are known to occur in fluid systems as well, through mechanisms that are understood to varying degrees \cite{CC1996,CC1996b,BV2012}. This remains an open area of research.

\subsection{Phase space measure and the Liouville theorem}
\label{sec:liouvillethm}

At the purely mathematical level, the statistical equilibrium approach is based on characterizing \emph{invariant} measures on the phase space $\Gamma$ of all possible functions $\omega({\bf r})$. Phase space integrals with respect to such a measure are therefore time independent and are used to construct physical equilibrium averages.

To be more specific, a probability density functional $\rho[\omega,t]$, which here assigns a positive real number to any given field realization $\omega = \{\omega({\bf r})\}_{{\bf r}\in D}$, evolves according to the conservation law
\begin{equation}
\partial_t \rho[\omega,t]
+ \nabla_\omega \cdot ({\bf V}[\omega] \rho[\omega]) = 0
\label{3.1}
\end{equation}
in which $\nabla_\omega$ is the (infinite dimensional) phase space gradient, and ${\bf V}[\omega]$ is the phase space velocity whose vector components are defined by each point ${\bf r} \in D$:
\begin{equation}
{\bf V}[\omega]({\bf r}) \equiv \partial_t \omega({\bf r})
= -{\bf v}[\omega]({\bf r}) \cdot \nabla \omega({\bf r}),
\label{3.2}
\end{equation}
derived from the equation of motion (\ref{2.4}). The linear functional ${\bf v}[\omega]({\bf r}) \equiv {\bf v}({\bf r})$ is given by the curl of (\ref{2.6}). The form (\ref{3.1}) ensures conservation of probability for any phase space volume co-moving with the phase space flow.

Now, an equilibrium probability density $\rho_\mathrm{eq}[\omega]$ allows one to define equilibrium averages in the form
\begin{equation}
\langle O[\omega] \rangle
= \int d\Gamma \rho_\mathrm{eq}[\omega] O[\omega].
\label{3.3}
\end{equation}
The functional integral is defined here by a limiting process
\begin{equation}
\int d\Gamma = \int D[\omega] \equiv \lim_{a \to 0}
\prod_i \int_{-\infty}^\infty d\omega({\bf r}_i)
\label{3.4}
\end{equation}
in which ${\bf r}_i$, $i = 1,2,3,\ldots,N_a = A_{\cal D}/a^2$, with $A_{\cal D}$ the area of ${\cal D}$, runs over a uniform grid (e.g., square lattice) with elements of area $a^2 \to 0$. For all such averages to be time-independent, $\rho_\mathrm{eq}$ must be as well and hence obey
\begin{equation}
\nabla_\omega \cdot ({\bf V}[\omega] \rho_\mathrm{eq}[\omega]) = 0.
\label{3.5}
\end{equation}

On the other hand, the equation of motion for any functional ${\cal I}[\omega,t]$, defined by
\begin{equation}
{\cal I}[\omega,t+dt] = {\cal I}[\omega + \partial_t \omega\, dt,t],
\label{3.6}
\end{equation}
takes the phase space advective form
\begin{equation}
\partial_t {\cal I}[\omega]
+ {\bf V}[\omega] \cdot \nabla_\omega {\cal I}[\omega] = 0.
\label{3.7}
\end{equation}
In particular, if ${\cal I}$ is a conserved integral then it must obey
\begin{equation}
{\bf V}[\omega] \cdot \nabla_\omega {\cal I}[\omega] = 0.
\label{3.8}
\end{equation}
The key observation is that if the phase space flow obeys the ``phase space incompressibility condition''
\begin{equation}
\nabla_\omega \cdot V[\omega] \equiv \int d{\bf r}
\frac{\delta V[\omega]({\bf r})}{\delta \omega({\bf r})} = 0,
\label{3.9}
\end{equation}
then the equilibrium measure condition (\ref{3.4}) reduces to
\begin{equation}
{\bf V}[\omega] \cdot \nabla_\omega {\bm \rho}_\mathrm{eq}[\omega] = 0.
\label{3.10}
\end{equation}
Comparing (\ref{3.8}), this corresponds to the requirement that $\rho_\mathrm{eq}$ \emph{be a conserved integral.} This is the content of the Liouville theorem.

\subsubsection{Liouville theorem for the Euler equation}
\label{subsec:liouvilleeuler}

The most transparent way to verify the Liouville theorem for the Euler equation, avoiding continuum functional derivatives, is to represent $\omega$ as a discrete orthogonal mode expansion on the finite domain ${\cal D}$. We consider an expansion of the stream function in Laplacian eigenmodes:
\begin{equation}
\psi({\bf r},t) = \sum_l \psi_l(t) \phi_l({\bf r})
\label{3.11}
\end{equation}
in which
\begin{equation}
-\nabla^2 \phi_l = \lambda_l \phi_l,
\label{3.12}
\end{equation}
with positive eigenvalues $\lambda_l > 0$ for a finite domain. The $\phi_l$ obey the same (Dirichlet or periodic) boundary conditions that $\psi$ does and may be taken to be real and orthonormal. It follows from (\ref{2.5}) that
\begin{eqnarray}
\omega({\bf r},t) &=& \sum_l \omega_l(t) \phi_l({\bf r}),\ \
\omega_l = \lambda_l \psi_l
\nonumber \\
{\bf v}({\bf r},t) &=& \sum_l \frac{1}{\lambda_l} \omega_l(t) {\bf v}_l({\bf r}),\ \
{\bf v}_l \equiv \nabla \times \phi_l,
\label{3.13}
\end{eqnarray}
and the equation of motion for $\omega_l$ may be derived in the form
\begin{equation}
\dot \omega_n(t) = \sum_{l,m} \frac{1}{\lambda_l}
W_{lmn} \omega_l(t) \omega_m(t)
\label{3.14}
\end{equation}
with coefficients
\begin{equation}
W_{lmn} = \int_{\cal D} d{\bf r} \phi_n({\bf r})
\nabla \phi_l({\bf r}) \times \nabla \phi_m({\bf r}).
\label{3.15}
\end{equation}
These are totally anisymmetric
\begin{equation}
W_{lmn}  = -W_{mln} = -W_{lnm}
\label{3.16}
\end{equation}
with the third one obtained via integration by parts, and making use of the free slip boundary condition to eliminate the boundary term. Using this representation, one obtains
\begin{eqnarray}
\nabla_\omega \cdot {\bf V}[\omega]
&=& \sum_n \frac{\partial \dot \omega_n}{\partial \omega_n}
\nonumber \\
&=& \sum_{m,n} W_{mnn} \omega_m
\left(\frac{1}{\lambda_m} - \frac{1}{\lambda_n} \right).
\label{3.17}
\end{eqnarray}
However, the coefficients $W_{mnn} = 0$ all vanish by virtue of the antisymmetry result. Thus, $\omega_n$ does not actually appear on the right hand side of (\ref{3.14}), trivially verifying the Liouville condition \cite{foot:realliouville}.

\subsection{Choice of statistical ensemble}
\label{sec:ensemblechoice}

\subsubsection{Microcanonical ensemble}
\label{subsec:microcanonical}

The choice of equilibrium measure goes by the name of \emph{statistical ensemble.} Perhaps the most transparent choice is the \emph{microcanonical} ensemble,
\begin{equation}
\rho_\mu[\omega;\varepsilon,{\bf c}] = \frac{1}{Z_\mu}
\delta(\varepsilon - E[\omega])
\prod_\gamma \delta(c_\gamma - C_\gamma[\omega])
\label{3.18}
\end{equation}
in which one constrains a particular value $c_\gamma$ to each conserved integral $C_\gamma[\omega]$ and we use the shorthand ${\bf c} = \{c_\gamma\}$. For the Euler equation this clearly involves an infinite product, which will be further characterized below. The energy is separated out explicitly for convenience. Equilibrium averages (\ref{3.3}) by construction limit the support of the phase space integral to vorticity fields constrained by the specified values $\varepsilon, {\bf c}$.

The partition function
\begin{equation}
Z_\mu(\varepsilon, {\bf c})
= \int D[\omega] \delta(\varepsilon - E[\omega])
\prod_\gamma \delta(c_\gamma - C_\gamma[\omega])
\label{3.19}
\end{equation}
serves to normalize $\rho_\mu$ as a probability density, but also defines the entropy function through the Boltzmann relation
\begin{equation}
S(\epsilon, {\bf c}) = \frac{1}{N_a} \ln[Z_\mu(\varepsilon, {\bf c})],
\label{3.20}
\end{equation}
in which the factor $1/N_a$ yields a finite, well defined result in the continuum limit (\ref{3.5}), here seen to play the role of the thermodynamic (infinite volume) limit in conventional systems. Explicit examples will be given below.

All thermodynamic quantities follow from the entropy function in the usual way. Most critically the inverse temperature
\begin{equation}
\beta \equiv \frac{1}{T} = \frac{\partial S}{\partial \varepsilon}
\label{3.21}
\end{equation}
is obtained from the energy derivative, and more generally the derivative
\begin{equation}
\mu_\gamma = T \frac{\partial S}{\partial c_\gamma}
\label{3.22}
\end{equation}
defines the thermodynamic field $\mu_\gamma$ conjugate to $c_\gamma$.

\subsubsection{Grand canonical ensemble}
\label{subsec:grandcanonical}

It is generally extremely difficult to compute delta function constrained integrals such as (\ref{3.19}). Instead one seeks to make use of the thermodynamic analogue of Lagrange multipliers by switching to a smoother probability distribution. Thus, the grand canonical ensemble is the defined by the Laplace transform
\begin{eqnarray}
\rho_\mathrm{GC}[\omega;\beta,{\bm \mu}] &=& \frac{Z_\mu}{Z_\mathrm{GC}}
\int d\varepsilon \int d{\bf c} e^{-\beta_a \left(\varepsilon
- \sum_\gamma \mu_\gamma c_\gamma \right)} \rho_\mu[\omega]
\nonumber \\
&=& \frac{1}{Z_\mathrm{GC}} e^{-\beta_a {\cal K}[\omega]}
\label{3.23}
\end{eqnarray}
with partition function
\begin{equation}
Z_\mathrm{GC}[\beta,{\bm \mu}] = \int D[\omega] e^{-\beta_a {\cal K}[\omega]}
\label{3.24}
\end{equation}
and statistical functional
\begin{equation}
{\cal K}[\omega] = E[\omega] -\sum_\gamma \mu_\gamma C_\gamma[\omega]
\label{3.25}
\end{equation}
now including fields ${\bm \mu} = \{\mu_\gamma\}$. These now replace the conserved integrals ${\bf c}$ as the fundamental thermodynamic variables. The subscript $a$ on $\beta$ allows for the fact that the inverse temperature $\beta_a = 1/T_a$ might need be scaled nontrivially in order to obtain a consistent thermodynamic description in the continuum limit. It will in fact be shown below that the scaling
\begin{equation}
\beta_a = \frac{\beta}{a^2} \equiv \frac{1}{T_a} = \frac{1}{Ta^2},
\label{3.26}
\end{equation}
is required, with finite values of $\beta = 1/T$ smoothly controlling the equilibrium state. This scaling is essentially required to control a nontrivial balance between energy and entropy (fluctuation) effects. Roughly speaking, equilibrium flows have lower temperature, $T_a = Ta^2 \to 0$, than that of any conventional thermodynamic system! The physical meaning of this will be discussed below.

\subsubsection{Thermodynamic free energy}
\label{subsec:freeenergy}

The partition function is now related to the thermodynamic free energy by
\begin{equation}
F(\beta,{\bm \mu}) = -\frac{1}{\beta_a}
\ln[Z_\mathrm{GC}(\beta,{\bm \mu})],
\label{3.27}
\end{equation}
and, with these scalings, is also finite and well defined in the continuum limit. Note that $\beta_a = (\beta/A_{\cal D}) N_a$ so that this actually involves the same $a$-scaling as the entropy (\ref{3.20}). From the definition (\ref{3.23}) the derivatives
\begin{eqnarray}
-\frac{\partial F}{\partial \mu_\gamma}
&=& \bar c_\gamma \equiv \langle C_\gamma[\omega] \rangle
\nonumber \\
\frac{\partial (\beta_a F)}{\partial \beta_a}
= \frac{\partial (\beta F)}{\partial \beta} &=& \bar K
\equiv \langle{\cal K}[\omega] \rangle
\nonumber \\
&=& \bar \varepsilon - \sum_\gamma \mu_\gamma \bar c_\gamma
\label{3.28}
\end{eqnarray}
produce the thermodynamic averages of the conserved integrals, defined here by
\begin{eqnarray}
\bar I \equiv \langle I[\omega] \rangle &=& \frac{1}{Z_\mathrm{GC}}
\int D[\omega] I[\omega] e^{-{\cal K}[\omega]}
\nonumber \\
&=& \frac{\int D[\omega] I[\omega] e^{-{\cal K}[\omega]}}
{\int D[\omega] e^{-{\cal K}[\omega]}}.
\label{3.29}
\end{eqnarray}
The standard equivalence of ensembles in the thermodynamic limit, which requires showing that the averages $\bar c_\gamma[\beta,{\bm \mu}]$ are in fact infinitely sharply peaked about a single unique value of $C_\gamma[\omega]$, can be shown to follow here from the continuum limit $N_a \to \infty$ (with peak width scaling as $1/\sqrt{N_a}$).

Other well known ensembles correspond to partial Laplace transforms over a subset of the conserved integrals. The canonical ensemble corresponds to transforming only the energy, resulting in statistical weight $e^{-\beta_a E[\omega]}$ multiplying the remaining delta functions. In a number of conventional systems the energy is actually the only conserved integral. There may be cases of ``ensemble inequivalence'' where dealing with the delta functions provides a more physically consistent approach \cite{BV2012}. However, even in such cases it is generally much simpler to apply the grand canonical approach and then use physical arguments to adapt it after the fact to more broadly enforce equivalence.

\subsubsection{Grand canonical formulation of the Euler equation}
\label{subsec:eulergc}

For the Euler equation the index $\gamma$ includes the continuous index $\sigma$ appearing in (\ref{2.14}), and one obtains the more explicit form
\begin{eqnarray}
\sum_\gamma \mu_\gamma C_\gamma[\omega]
&\to& \int d\sigma \mu(\sigma) g[\omega;\sigma] + \mu_P P[\omega]
\nonumber \\
&=& \int_{\cal D} d{\bf r} \{\mu[\omega({\bf r})]
+ \mu_P \alpha({\bf r}) \omega({\bf r}) \} \ \ \ \ \ \
\label{3.30}
\end{eqnarray}
in which the conserved momentum (\ref{2.12}), when it exists, enters with conjugate field $\mu_P$. The 1D field function $\mu(\sigma)$ is conjugate to the conserved function $g(\sigma) \equiv g[\omega;\sigma]$, promoted here to a functional of the vorticity. Inserting this form into (\ref{3.25}) produces the field theory displayed in the first row of Table \ref{tab:fields}.

\section{Thermodynamics of the Euler equation: Exact solution}
\label{sec:tdeulerexact}

We will now show, quite remarkably, that the Euler equation free energy (\ref{3.27}) may be computed \emph{exactly.} More specifically, the evaluation of the full phase space integral (\ref{3.24}) may be reduced to a variational equation for the free energy from which the equilibrium vorticity function
\begin{equation}
\omega_0({\bf r};\beta,{\bm \mu}) = \langle \omega({\bf r}) \rangle
\label{4.1}
\end{equation}
is obtained as a solution to a (highly nonlinear) PDE generated by the corresponding Euler--Lagrange equation. The derivation here will be physically motivated rather than rigorous---full details may be found in Ref.\ \cite{MWC1992}. Such variational approaches often emerge as approximate ``mean field'' descriptions of conventional thermodynamic systems. Here the mean field form is in fact exact due to the long range (Coulomb-like) interactions (\ref{2.9}) between vortices.

\subsection{Mean field approach}
\label{sec:mfapproach}

The key property of the energy function (\ref{2.9}) is that it is dominated by the long range nature of $G({\bf r},{\bf r}')$: in the macroscopic coherent flow regime of interest here the stream function (and therefore the advection velocity field) is dominated by the global integral (\ref{2.6}) over the entire domain. In contrast to systems with local interactions, the contribution from a small area $l^2$ about ${\bf r}$ here scales as $l^2 \ln(l) \to 0$. It follows that if one considers a fluctuation $\omega({\bf r}) - \omega_0({\bf r})$ about the equilibrium field one may accurately replace
\begin{eqnarray}
E[\omega] &\to&  E[\omega_0] + \int_{\cal D} d{\bf r}
\left[\frac{\delta E}{\delta \omega({\bf r})} \right]_{\omega = \omega_0}
[\omega({\bf r}) - \omega_0({\bf r})]
\nonumber \\
&=& E[\omega_0] + \int_{\cal D} d{\bf r} \psi_0({\bf r})
[\omega({\bf r}) - \omega_0({\bf r})]
\nonumber \\
&=& -E[\omega_0]  + \int_{\cal D} d{\bf r} \psi_0({\bf r}) \omega({\bf r})
\label{4.2}
\end{eqnarray}
in which
\begin{equation}
\psi_0({\bf r};\beta,{\bm \mu}) \equiv \langle \psi({\bf r}) \rangle
= \int d{\bf r}' G({\bf r},{\bf r}') \omega_0({\bf r};\beta,{\bm \mu})
\label{4.3}
\end{equation}
is the equilibrium stream function. The key observation asymptotically.

The inverse relationship
\begin{equation}
\omega_0 = -\nabla^2 \psi_0.
\label{4.4}
\end{equation}
then also follows. In conventional particle systems $G$ is typically a short ranged microscale interaction, $\psi_0$ is therefore dominated by local fluctuations on the same scale as $\omega$, and (\ref{4.2}) is at best approximate. Here the Casimirs strongly bound the fluctuations of $\omega$, the Green function effectively performs a self-averaging operation so that $\psi({\bf r}) - \psi_0({\bf r}) \to 0$ in the continuum limit with probability one, and (\ref{4.2}) becomes exact \cite{foot:ablim}.

The form of $\psi_0$ must now be determined self consistently by using (\ref{4.2}) to compute the free energy. Substituting (\ref{4.2}) and (\ref{3.30}) into (\ref{3.25}) one obtains
\begin{equation}
{\cal K}[\omega] \to -E[\omega_0] + \int d{\bf r}
K[\psi_0({\bf r}) - \mu_P \alpha({\bf r}),\omega({\bf r})]
\label{4.5}
\end{equation}
with 2D function
\begin{equation}
K(\tau,\sigma) = \sigma \tau - \mu(\sigma).
\label{4.6}
\end{equation}
This form is now purely local in the fluctuating field $\omega({\bf r})$. The temperature scaling (\ref{3.26}) is now seen to be chosen to enable the replacement
\begin{equation}
\beta_a \int d{\bf r} \to \beta_a a^2 \sum_i = \beta \sum_i,
\label{4.7}
\end{equation}
and the partition function then follows in the product form
\begin{eqnarray}
Z_\mathrm{GC} &=& \lim_{a \to 0} e^{\beta_a E[\omega_0]}
\prod_i \int_{-\infty}^\infty
d\omega_i e^{-\beta K[\psi_0({\bf r}_i) - \mu_P \alpha({\bf r}_i),\omega_i]}
\nonumber \\
&=& \lim_{a \to 0}  e^{\beta_a E[\omega_0]}
e^{-\beta \sum_i W[\psi_{0,i}- \mu_P \alpha({\bf r}_i)]}
\label{4.8}
\end{eqnarray}
in which we define the 1D function
\begin{equation}
W(\tau) = -\frac{1}{\beta}
\ln\left[\int d\sigma e^{-\beta  K(\tau,\sigma)} \right],
\label{4.9}
\end{equation}
essentially the Laplace transform of $e^{\beta \mu(\sigma)}$. Explicit forms for $W$ obtained from simple model forms for $\mu(\sigma)$ will be discussed below.

Taking the logarithm of (\ref{4.8}) and restoring continuum notation, the final free energy functional (\ref{3.27}) takes the form
\begin{eqnarray}
&&{\cal F}[\psi_0;\beta,{\bm \mu}]
= -E[\omega_0] + \int_{\cal D} d{\bf r}
W[\psi_0({\bf r}) - \mu_P \alpha({\bf r});\beta,{\bm \mu}]
\nonumber \\
&&\ \ \ \ =\ \int_{\cal D} d{\bf r}\left\{-\frac{1}{2} |\nabla \psi_0|^2
+ W[\psi_0({\bf r}) - \mu_P \alpha({\bf r});\beta,{\bm \mu}]\right\}
\nonumber \\
\label{4.10}
\end{eqnarray}
with the dependence on the thermodynamic fields $\beta,{\bm \mu}$ now highlighted explicitly. The scaling (\ref{3.26}) is again confirmed to yield a well defined finite result.

\begin{figure}

\includegraphics[width=3.2in,viewport = 275 110 670 470,clip]{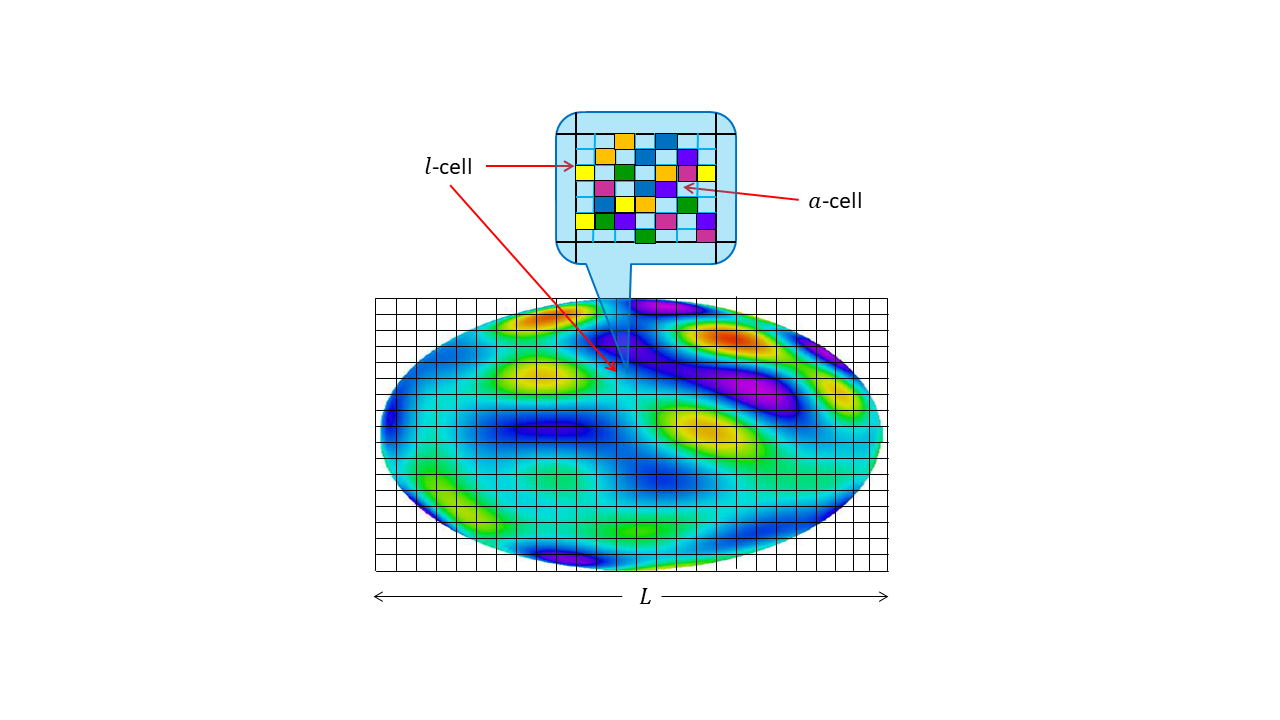}

\caption{Illustration of separation of scales entering the exact thermodynamic solution. The vortex self-advection is dominated by the large-scale flow, while the small scale fluctuations asymptotically obey a simple $a$-cell permutation rule generating the microscale entropy (\ref{4.19}) characterizing each intermediate scale $l$-cell. Within each $l$-cell one may define the local vorticity distribution $n_0({\bf r}_l,\sigma)$ which has a well defined continuum limit $a,l \to 0$ but in such a way that $l/a \to \infty$. Its first moment defines the equilibrium vorticity (\ref{4.12}) and its area integral is constrained by the Casimir function (\ref{4.13}). This illustrates the formal limiting process by which, e.g., a discrete set of ($a$-scale) vorticity levels controlled by the Casimirs produces a smooth ($l$-scale) average.}

\label{fig:micromacro}
\end{figure}

A self consistent equation for $\psi_0$ is obtained by generalizing the free energy calculation to compute equilibrium averages (\ref{3.29}). The fundamental quantity needed is the vorticity distribution function (illustrated in Fig.\ \ref{fig:micromacro})
\begin{eqnarray}
n_0({\bf r},\sigma) &=& \langle \delta[\sigma - \omega({\bf r})] \rangle
\label{4.11} \\
&\to& e^{\beta W[\psi_0({\bf r}) - \mu_P \alpha({\bf r})]}
e^{-\beta K[\psi_0({\bf r}) - \mu_P \alpha({\bf r}),\sigma]}.
\nonumber
\end{eqnarray}
This simple result follows from the cancelation of the integrals over all other $\omega_i \neq \omega({\bf r})$ between the numerator and denominator of (\ref{3.29}). This function quantifies the fluctuations of the vortex field in the microscopic neighborhood of any given point ${\bf r}$ (defined by the $l$-cells in Fig.\ \ref{fig:micromacro}). In particular, the mean vorticity is derived in the form
\begin{eqnarray}
\omega_0({\bf r}) &=& -\nabla^2 \psi_0({\bf r})
\nonumber \\
&=& \int d\sigma \sigma n_0({\bf r},\sigma)
\nonumber \\
&=& \frac{\int \sigma d\sigma
e^{-\beta K[\psi_0({\bf r}) - \mu_P \alpha({\bf r}),\sigma]}}
{\int d\sigma e^{-\beta K[\psi_0({\bf r}) - \mu_P \alpha({\bf r}),\sigma]}}.
\label{4.12}
\end{eqnarray}
The right hand side is a local function of $\psi_0({\bf r})$, so that we have produced a type of nonlinear Poisson equation for $\psi_0$.

In addition, the Casimirs (\ref{2.14}) are recovered from the area integral
\begin{equation}
g(\sigma) = \int_{\cal D} d{\bf r} n_0({\bf r},\sigma),
\label{4.13}
\end{equation}
which allows one, in principle, to invert for $\mu(\sigma)$ for specified $g(\sigma)$. The identical result may be shown to follow from the functional derivative
\begin{equation}
g(\sigma) = -\frac{\delta {\cal F}}{\delta \mu(\sigma)}.
\label{4.14}
\end{equation}
This derivative is performed only with respect to the explicit ${\bm \mu}$ dependence in (\ref{4.10}), keeping $\psi_0$ fixed. This works because the self-consistency condition (\ref{4.12}) is equivalent to the extremum condition
\begin{equation}
\left(\frac{\delta{\cal F}[\psi]}{\delta \psi({\bf r})} \right)_{\psi = \psi_0} = 0
\label{4.15}
\end{equation}
which zeros out the $\delta \psi_0/\delta \mu(\sigma)$ contribution to (\ref{4.14}).

The equilibrium momentum
\begin{equation}
\langle P[\omega] \rangle = \int d{\bf r} \alpha({\bf r}) \omega_0({\bf r})
= - \frac{\partial {\cal F}}{\partial \mu_P}
\label{4.16}
\end{equation}
may similarly derived either from $\omega_0$ or from the free energy derivative. The mean fluid kinetic energy
\begin{equation}
\langle E[\omega] \rangle
= \frac{1}{2} \int_{\cal D} d{\bf r} |{\bf v}_0({\bf r})|^2
= \frac{1}{2} \int_{\cal D} d{\bf r} |\nabla \psi_0({\bf r})|^2
\label{4.17}
\end{equation}
follows as well either by substituting $\omega_0({\bf r}) = \nabla \times {\bf v}_0({\bf r})$ into (\ref{2.9}) or from the $\beta$ derivative exhibited in (\ref{3.28}).

\subsection{Microscale entropy}
\label{sec:microentropy}

The distribution function (\ref{4.11}) also allows one to introduce the important concept of the microscale fluid entropy. The equilibrium flow defined by $\omega_0$ and $\psi_0$ is smooth, in general infinitely differentiable on any finite physical length scale. The equilibration process may be thought of as the completion of the inverse cascade of energy, which serves to create the inhomogeneous flow on the domain scale $A_{\cal D}$, and the forward cascade of enstrophy (and all other Casimirs) to infinitesimal scales that render the microscale fluctuations invisible. Of course, additional physical dissipation processes such as viscosity will eventually smooth out these microscales, but this not necessary to make sense of the idealized fluid equilibria considered here.

Using (\ref{4.11}) the equilibrium entropy
\begin{equation}
S = -\frac{\partial {\cal F}}{\partial T}
= \beta^2 \frac{\partial {\cal F}}{\partial \beta}
\label{4.18}
\end{equation}
may be expressed in the classic information theoretic form
\begin{equation}
S[n_0] = -\int_{\cal D} d{\bf r}
\int d\sigma n_0({\bf r},\sigma) \ln[n_0({\bf r},\sigma)].
\label{4.19}
\end{equation}
This precisely captures the information lost in going from the exact microscale specification of the finely mixed vorticity field (Fig.\ \ref{fig:micromacro}) at any given instant of time to the time-independent equilibrium average, in which only $\omega_0$ is specified.

For any given distribution $n_0$, not necessarily equilibrium, one may derive (\ref{4.19}) from the Boltzmann formula
\begin{equation}
S[n_0] = \frac{1}{N_a} \ln\{N[n_0]\},
\label{4.20}
\end{equation}
which may be compared to the microcanonical expression (\ref{3.20}). The derivation proceeds via the previously described limiting process in which one counts the total number of ways $N[n_0]$ to distribute the $(l/a)^2$ vorticity levels contained in the intermediate scale area $l^2$, with level populations constrained by $n_0$ (essentially an $a$-cell permutation count repeated over all $l$-cells). In fact, an alternative rigorous microcanonical approach to deriving the free energy functional (\ref{4.10}) is to maximize $S[n_0]$ subject to the all of the conserved integral constraints \cite{MWC1992}. The maximal solution for $n_0$ is recovered precisely in the form (\ref{4.11}).

\subsection{Rotating fluids and generalization to the beta plane}
\label{sec:rotbetaplane}

Before turning to explicit examples and further generalization of the theory, it is worth treating the simplest extension to rotating fluids. The beta plane approximation incorporates planetary rotation through the generalization
\begin{equation}
\frac{D {\bf v}}{Dt} + f({\bf r}) \hat {\bf z} \times {\bf v} = -\nabla p
\label{4.21}
\end{equation}
in which $f({\bf r}) = 2 \hat {\bf z} \cdot {\bm \Omega} = 2 \Omega \sin(\theta_L)$ is the Coriolis function derived from the local vertical projection of the angular rotation vector ${\bm \Omega}$ corresponding to latitude $\theta_L({\bf r})$. The curl of this equation leads to self-advection of the potential vorticity
\begin{equation}
\frac{D \omega_P}{Dt} = 0,\ \
\omega_P({\bf r}) = \omega({\bf r}) + f({\bf r})
\label{4.22}
\end{equation}
exhibiting to the sum of local and frame of reference rotation rates. The kinetic energy (\ref{2.9}) remains unchanged, but is now expressed in terms of $\omega_P$ by substituting $\omega = \omega_P - f$. Similarly, for the momenta, which are now conserved only if $f({\bf r})$ possesses the required invariance---constant latitude (east-west) periodic strip, or disc or annulus surrounding the pole.

The equilibrium free energy follows in a form identical to (\ref{4.10}), but with
\begin{equation}
\psi_0({\bf r}) \to \psi_P({\bf r}) - F({\bf r}),
\label{4.23}
\end{equation}
in which $F$ is the solution to Poisson equation
\begin{equation}
-\nabla^2 F({\bf r}) = f({\bf r}).
\label{4.24}
\end{equation}
For linear $f = \beta y$ on a strip, or $f = \beta r$ on a disc or annulus (beta plane linear approximation), one obtains the cubic form $F = -\frac{1}{6} \beta y^3$ or $F = -\frac{1}{9} \beta r^3$.  The result is the combination
\begin{equation}
\Psi_P({\bf r}) = F({\bf r}) + \mu_P \alpha({\bf r})
\label{4.25}
\end{equation}
acting as an ``external potential'' $\psi_0 - \Psi_P$ inside the $W$ function in (\ref{4.10}). Since the two functional forms are different [linear or quadratic---see (\ref{2.12})---vs.\ cubic], the result can be an interesting balance or competition between angular momentum and Coriolis effects. Such effects can stabilize large scale vortex structures, such as Jupiter's Red Spot, at a particular latitude, e.g., near a local extremum of $\Psi_P$ \cite{BV2012,MWC1992,MR1994}.

\subsubsection{More general curvilinear domains}
\label{subsec:curvedsurfgen}

More generally the Euler equation on a 2D curved (in particular spherical) surface, with and without rotation, may developed as well \cite{DQM2015}. The vorticity and stream function may be defined by adopting appropriate curvilinear coordinates, and the generalization of the self advection dynamics (\ref{2.4}) for the vorticity then follows. The conserved Casimir area integrals then follow immediately as well, as does the statistical theory leading to a free energy functional in a form very similar to (\ref{4.10}).

\begin{figure}

\includegraphics[width=3.2in,viewport = 200 60 720 440,clip]{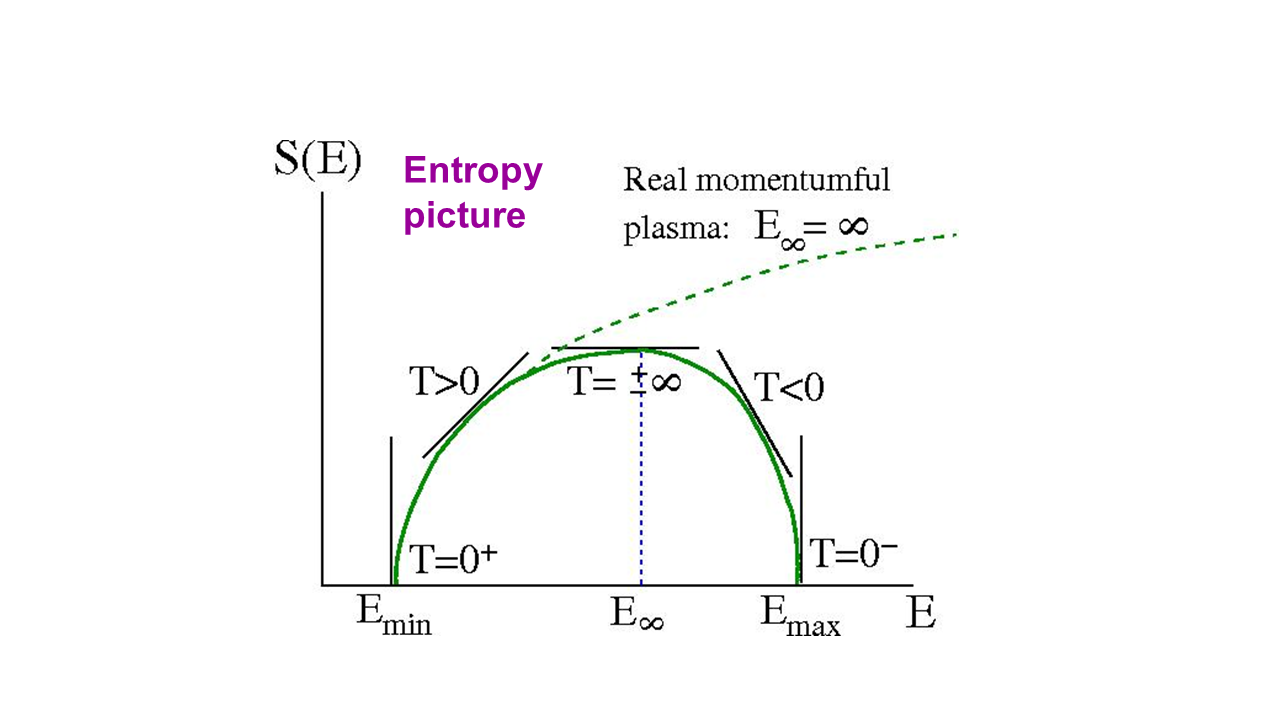}

\caption{Schematic illustration of the entropy function $S(E)$ associated with the two level system (\ref{4.26}), and also of the point vortex system pictured in Fig.\ \ref{fig:ptvortex}. As described in the text, the Casimir constraints on the vorticity allow for both positive and negative temperatures, and corresponding entropy limited to a finite energy interval, vanishing with infinite slope at both ends. This general picture will hold for any $g(\sigma)$ with bounded support. The dashed line corresponds to conventional particle systems in which the momentum degree of freedom can absorb unbounded energy.}

\label{fig:entropy}
\end{figure}

\subsection{Simplified model examples}
\label{sec:simplemodel}

\begin{figure*}

\includegraphics[width=2.3in,viewport = 30 0 380 300,clip]{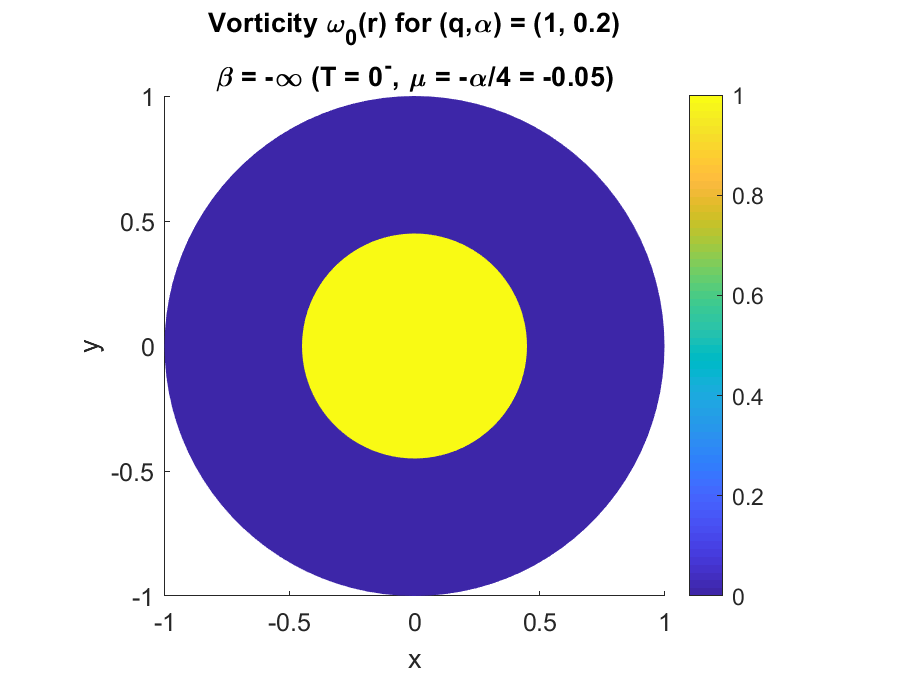}
\includegraphics[width=2.3in,viewport = 30 0 380 300,clip]{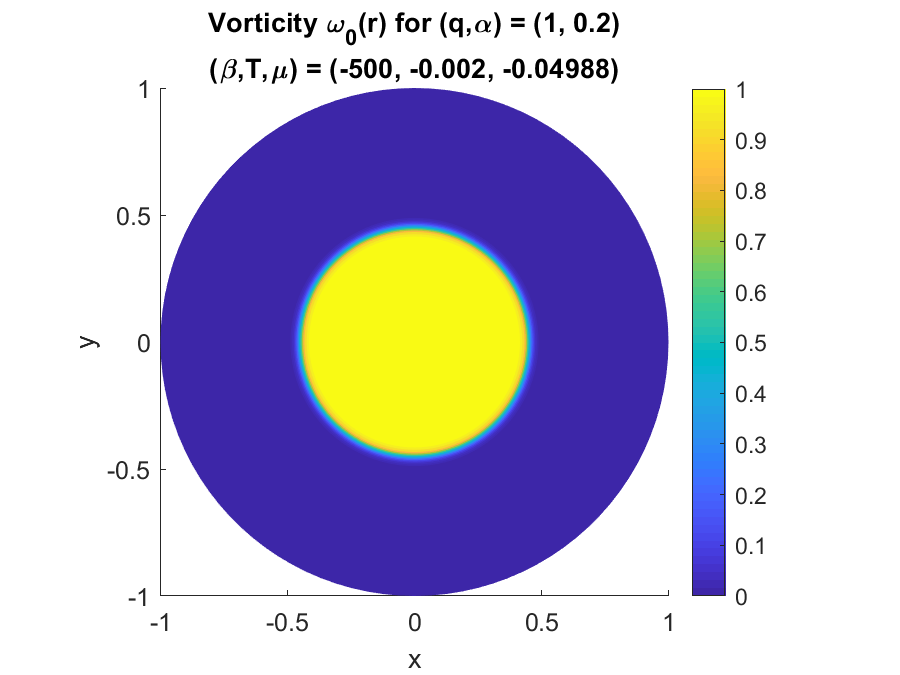}
\includegraphics[width=2.3in,viewport = 30 0 380 300,clip]{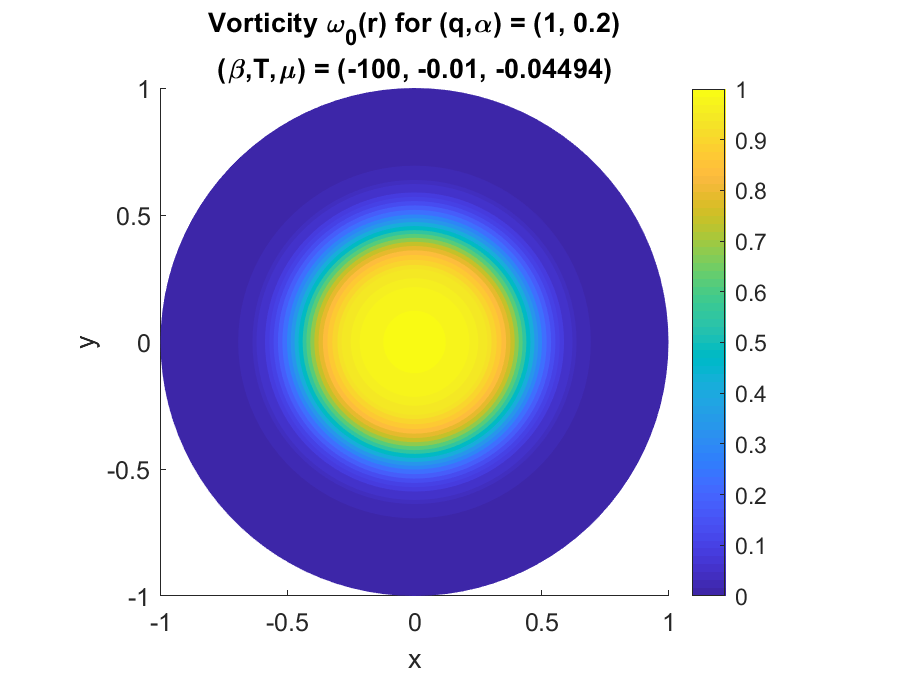}

\includegraphics[width=2.3in,viewport = 30 0 380 300,clip]{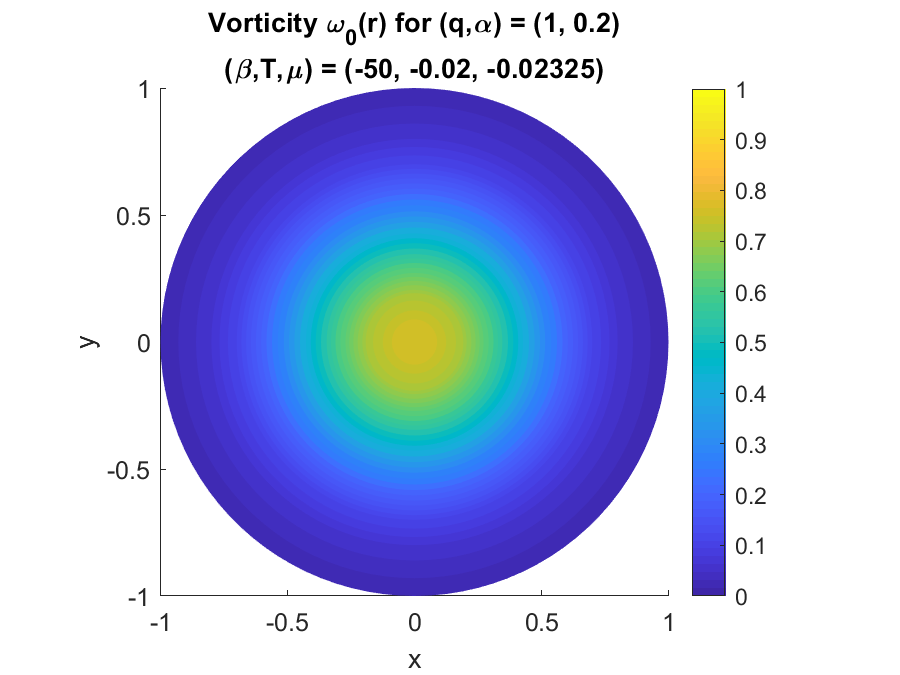}
\includegraphics[width=2.3in,viewport = 30 0 380 300,clip]{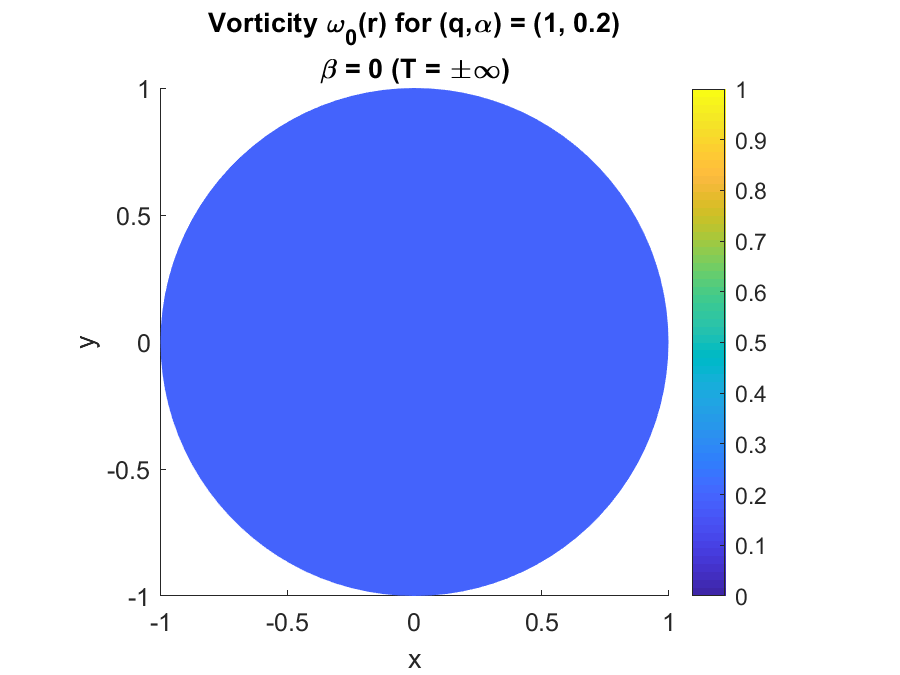}
\includegraphics[width=2.3in,viewport = 30 0 380 300,clip]{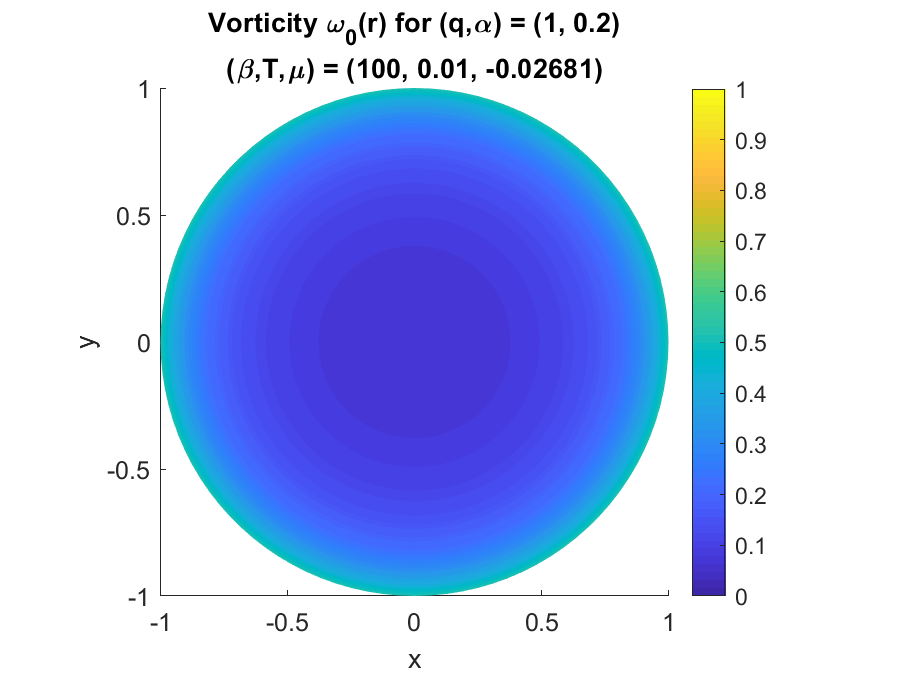}

\includegraphics[width=2.3in,viewport = 30 0 380 300,clip]{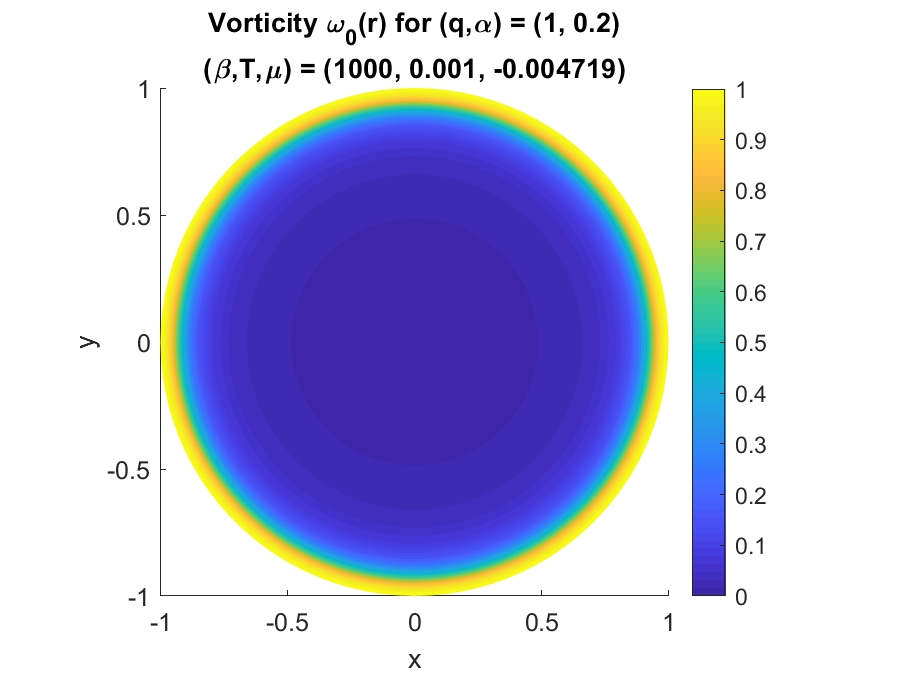}
\includegraphics[width=2.3in,viewport = 30 0 380 300,clip]{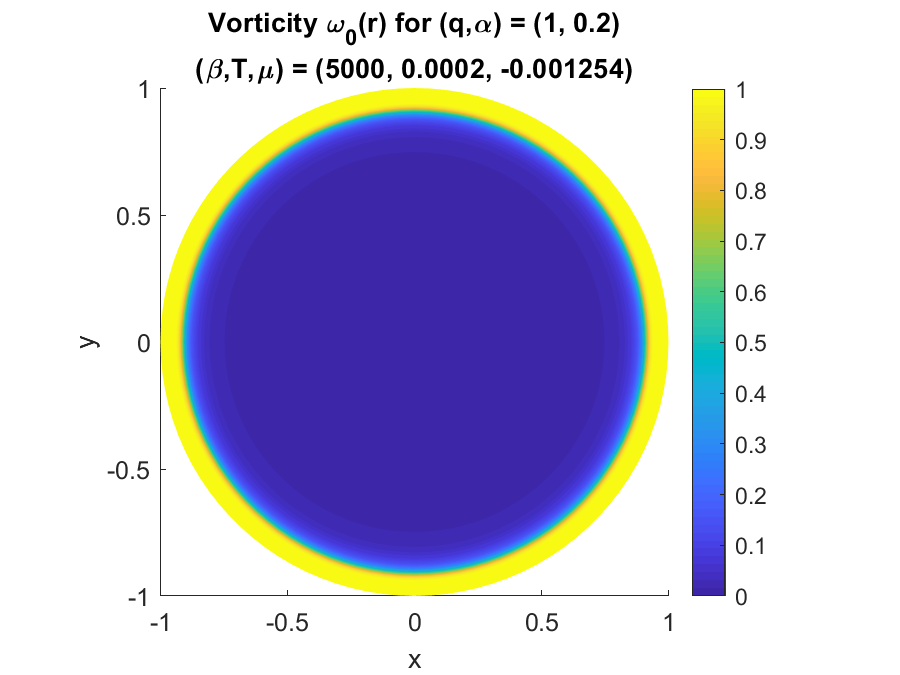}
\includegraphics[width=2.3in,viewport = 30 0 380 300,clip]{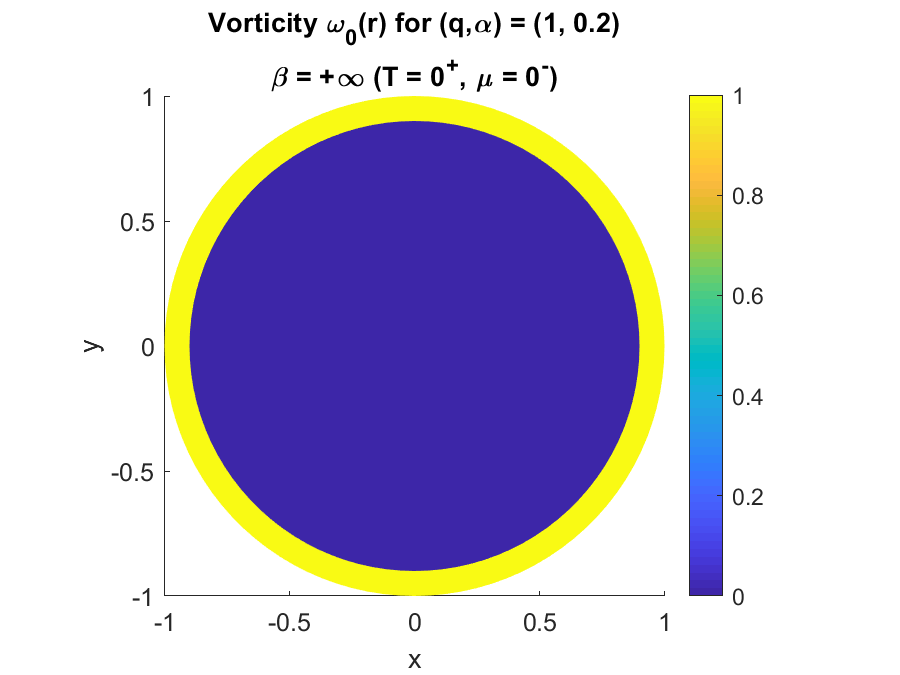}

\caption{Example equilibrium vorticity profiles $\omega_0({\bf r})$ for the two level system on the unit disk for a sequence of inverse temperatures $-\infty \leq \beta \leq \infty$, obtained by numerically solving the nonlinear Laplace equation (\ref{4.28}). Vorticity level $q = 1$ occupies fractional area $\alpha = 0.2$, hence total vorticity $\Omega_0 = \pi\alpha$. For each temperature, the Lagrange multiplier $\mu_q(\beta)$ must be determined iteratively to satisfy this constraint. As seen the $\beta = -\infty$ ($T = 0^-$) the maximum energy solution gathers all vorticity near the disc center, while the $\beta = +\infty$ ($T = 0^+$) solution compacts all vorticity against the disc boundary. The $\beta = 0$ ($T = \pm \infty$) maximum entropy solution distributes the vorticity uniformly (center panel).}

\label{fig:2leveleg}
\end{figure*}

The equilibrium equation (\ref{4.12}) looks quite complicated, but some very interesting, physically meaningful results may be derived by specializing to few parameter models. We will focus on the two-level system
\begin{equation}
g(\sigma) = A_0 \delta(\sigma) + A_q \delta(\sigma - q),\ \
A_0 + A_q = A_{\cal D}
\label{4.26}
\end{equation}
in which the vorticity field is constrained to take values 0 or $q$ only (illustrated in Fig.\ \ref{fig:turbmix}). Since for given domain area $A_{\cal D}$ there is only a single degree of freedom, one may normalize
\begin{equation}
e^{\beta \mu(\sigma)} = \delta(\sigma) + e^{\beta \mu_q} \delta(\sigma - q),
\label{4.27}
\end{equation}
in which the single conjugate field $\mu_q$ is used to adjust the relative areas of the vortex ``charges.'' Substituting into (\ref{4.12}) one obtains the equilibrium equation
\begin{equation}
\omega_0({\bf r}) = -\nabla^2 \psi_0({\bf r})
= \frac{q}{e^{\beta [q \psi_0({\bf r}) - \mu_q]} + 1}
\label{4.28}
\end{equation}
with a Fermi-like distribution function on the right hand side, and in which for simplicity we set the momentum to zero (if it exists) by taking $\mu_P = 0$. For large $\beta \to \infty$ ($T \to 0^+$) the solution is $\omega_0 = 0$ on the region where $\psi_0 < \mu_q/q$, and $\omega_0 = q$ on the compliment, so that the equilibrium  solution is also two-level. This solution corresponds to the lowest possible energy state, and by Gauss's law spreads the vorticity out as much as possible (equal-signed charges repel), distributing it up against the boundary of ${\cal D}$. On the other hand, for large $\beta \to -\infty$ ($T \to 0^-$), which is perfectly allowed in this system, the two regions switch roles, with $\omega_0 = q$ on the region where $\psi_0 > \mu_q/q$ and $\omega_0 = 0$ on the compliment. The solution corresponds to the highest possible energy (equal-signed charges now effectively attract), and the result is a single compact vortex somewhere in the interior of ${\cal D}$. Varying $\mu_q$ varies the position of the vortex boundary, hence size of the vortex. As one varies $-\infty < \beta < \infty$ the vortex edge will be smeared out on the scale $|T| = 1/|\beta|$ and the solution will continuously interpolate between these two extremes. Figure \ref{fig:2leveleg} illustrates these results for a unit disc domain. The solutions for this simple case are azimuthally symmetric, functions of the radius $r$ alone.

This behavior of the solution as one varies $-\infty < \beta  = \partial S/\partial E < \infty$ in accompanied by a very interesting picture of the energy dependence of the entropy $S(E)$, illustrated in Fig.\ \ref{fig:entropy}. The entropy vanishes for $|\beta| \to \infty$  ($T \to 0^\pm$) corresponding to the minimum energy $E_\mathrm{min}$ (vorticity compacted against the boundary) and maximum energy $E_\mathrm{max}$ (vorticity compacted at the center). The maximum entropy occurs for $\beta = 0$ (maximally disordered uniform vorticity state at $T \to \pm \infty$) but at some intermediate value of the energy. For conventional particle systems, the particle momenta are permitted to grow without bound and $S(E)$ diverges with $E \to \infty$---the curve never turns over and negative temperatures are forbidden.

More interesting behaviors may observed in annular domains \cite{MWC1992,CC1996,Marcus1988,Marcus1990} where the azimuthal symmetry may be broken (a form of second order phase transition). Dynamically, a zonal jet (symmetric vortex ring in this case) becomes unstable and at late time forms a simply connected Red Spot-like vortex. Within the equilibrium theory, the energy advantage of a more compact shape leads to spontaneous azimuthal symmetry breaking for decreasing negative temperature.

Other interesting behaviors may be explored using the three level system
\begin{eqnarray}
g(\sigma) &=& A_0 \delta(\sigma)
+ A_q [\delta(\sigma - q) + \delta(\sigma + q)]
\nonumber \\
e^{\beta \mu(\sigma)} &=& \delta(\sigma)
+ e^{\beta \mu_q} [\delta(\sigma - q) + \delta(\sigma + q)]
\label{4.29}
\end{eqnarray}
with $A_0 + 2A_q = A_{\cal D}$, and equilibrium equation
\begin{equation}
\omega_0 = -\nabla^2 \psi_0 = -q \frac{\sinh(\beta q \psi_0)}
{\cosh(\beta q \psi_0) + \frac{1}{2} e^{-\beta \mu_q}}.
\label{4.30}
\end{equation}
The model here is simplified by enforcing symmetry between charges $\pm q$. High energy, negative temperature equilibria, for example, with two separated, opposite-signed vortex blobs may be constructed. On the other hand, low energy states correspond to fine-scale intermixing of the two charges, generating a (conventional) featureless, neutral system with no macroscale flow structure.

Breaking the symmetry between the charges, $\mu_q \neq \mu_{-q}$, allows one to separately control the relative size of these blobs, and eliminate full cancelation at positive temperatures.

\section{Some limitations of the statistical equilibrium hypothesis}
\label{sec:eqhyplimits}

\subsection{Metastable steady states}
\label{sec:metastable}

Ergocity is a statement that long time averages are equivalent to an equilibrium average, namely that any initial condition $\omega({\bf r},0)$ will explore essentially all of the phase space permitted by the basic conservation laws. There are a number of cases where this assumption can fail, and the highly constrained nature of 2D flows exacerbates this (see Sec.\ 3.1.2 of Ref.\ \cite{BV2012} for some discussion on this point). This is in contrast, for example, to conventional gases with their randomly moving atoms and molecules with their interpenetrating motion trajectories. Thus, equilibration of 2D flows is perhaps more closely analogous to dense, glassy system dynamics with strong barriers to individual particle motions.

Some of these barriers can actually be understood as local rather than global minima of the free energy functional. Examples include separated compact eddies that orbit each other, failing to merge (as would be entropically favored) above a critical separation \cite{CC1996}. There are numerous examples of more conventional systems that show analogous behaviors, including decay of superflow (which requires, e.g., nucleation of an eventually system-spanning vortex ring), and metastability of certain crystal structures such as that of carbon's diamond.  Detailed 2D Euler equation numerical simulations also show evidence for different levels of equilibration in different spatial regions, depending on the strength of local mixing dynamics \cite{CC1996b}.

\subsection{Viscosity effects}
\label{sec:visceff}

The varying roles of microscale viscosity should also be pointed out. Of course, viscous effects standing in for thermal exchange between the micro- and macro-levels (as well as bottom and other forms of friction with the world outside the idealized 2D domain), will eventually lead to strong violations of the vast majority of conservation laws, producing decay to an essentially trivial flow. The statistical mechanics approach can at best be valid on intermediate time scales where such effects can be ignored. 

At a more subtle level, the $l$-cell picture in Fig.\ \ref{fig:micromacro} will be the first to be violated. The individual $a$-cells will diffusively mix to form a uniform, average vorticity essentially coinciding with the local average $\omega_0({\bf r})$ defined by the second line of (\ref{4.12}). It is important to emphasize that this by itself does not violate the statistical mechanics predictions: the macro-scale flow is insensitive to such microscale averaging. Although the Casimir function $g(\sigma)$ is lost in this process, replaced by the ``diffusion-mixed'' form 
\begin{equation}
g_d(\sigma) = \int_{\cal D} d{\bf r} \delta[\sigma - \omega_0({\bf r})], 
\label{5.1}
\end{equation}
it can be shown \cite{foot:dvc} that $\omega_0({\bf r})$ can be consistently derived from $g_d(\sigma)$ as the \emph{extremum energy} solution of the equilibrium equation (\ref{4.12}) ($|T| \to 0$ or $|\beta| \to \infty$, depending the sign of $\beta$ prior to the action of the viscosity).

A more nuanced definition of the intermediate time scale is therefore that it be small enough that large scale flows are not significantly affected by viscosity, but not so small that it unnecessarily forbids the occurrence of simultaneous fine scale diffusive mixing during the course of the late time turbulent cascade.

\subsection{Strongly fluctuating long-lived states}
\label{sec:strongfluct}

Finally we note that equilibration dynamics can be strongly affected by the topology of the 2D domain. Thus, equilibration on the surface of a sphere (rather than in a flat bounded or doubly periodic planes) is found to fail much more catastrophically, with a macroscopically fluctuating chaotic vorticity field surviving for all achievable computation times \cite{QM2014,DQM2015}. Conservation of the full angular momentum vector in the spherical geometry (rather than just a vertical component) ensures that the vorticity cannot condense into a dipole pattern if the initial state has zero total angular momentum. Numerical experiments show that most (but not all) of the vorticity condenses instead into a quadrupole with two positive vortices and two negative vortices; small but satellite vortices also persist \cite{DQM2015}. The quadrupolar configuration oscillates, likely chaotically, at long times, consistent with the dynamics of the much simpler problem of four point vortices on the surface of sphere \cite{D2020}.

\section{General statistical theory of single-field systems}
\label{sec:statgen1d}

We next consider generalization of the statistical approach to other 2D systems characterized by an infinite number of conserved Casimir-type integrals constraining the dynamics of a single scalar field \cite{W2006}. The former requires the existence of a self-advecting field $q({\bf r},t)$ with equation of motion
\begin{equation}
\frac{Dq}{Dt} \equiv \partial_t q + {\bf v} \cdot \nabla q = 0
\label{6.1}
\end{equation}
generalizing (\ref{2.4}) in the sense that the relation between $q$ and the velocity field ${\bf v}$ may be more general. A convenient way to constrain this relationship, and simultaneously ensure existence of a consistent statistical theory, is demand that (\ref{6.1}) be derived from a Hamiltonian equation of motion
\begin{equation}
\partial_t q = \{q, E[q]\}
\label{6.2}
\end{equation}
with (conserved) energy function $E[q]$ and Poisson bracket of two functionals $A[q]$, $B[q]$ defined by \cite{HMRW1985}
\begin{equation}
\{A[q],B[q]\} = \int d{\bf r} q({\bf r})
\nabla \frac{\delta A}{\delta q({\bf r})}
\times \nabla \frac{\delta B}{\delta q({\bf r})}.
\label{6.3}
\end{equation}
In particular, if one defines the stream function by
\begin{equation}
\psi({\bf r}) = \frac{\delta E}{\delta q({\bf r})}
\label{6.4}
\end{equation}
then (\ref{6.2}) takes the form
\begin{equation}
\partial_t q = \nabla \psi \times \nabla q
= -(\nabla \times \psi) \cdot \nabla q
\label{6.5}
\end{equation}
which is exactly (\ref{6.1}) with the usual stream function relation (\ref{2.3}), and from which incompressibility of ${\bf v}$ also follows immediately. Conservation of the Casimirs (\ref{2.13})--(\ref{2.15}), with $q$ replacing $\omega$, follows immediately as well.

\subsection{Statistical mechanics}
\label{sec:statmechgen}

The Liouville theorem
\begin{equation}
\int d{\bf r} \frac{\delta \dot q({\bf r})}{\delta q({\bf r})} = 0
\label{6.6}
\end{equation}
follows directly from the Hamiltonian structure. Specifically, using a mode representation (\ref{3.11}) for $q$, one obtains
\begin{equation}
\dot q_l = \sum_{m,n} W_{lmn} \frac{\partial E}{\partial q_m}
\label{6.7}
\end{equation}
with coefficients defined by (\ref{3.15}), and it follows that
\begin{equation}
\sum_l \frac{\partial \dot q_l}{\partial q_l}
= \sum_{l,m,n} W_{lmn} \frac{\partial^2 E}{\partial q_l \partial q_m} = 0.
\label{6.8}
\end{equation}
in which the vanishing follows because $W_{lmn}$ is antisymmetric in $l,m$ [see (\ref{3.16})] while the mixed partial is even. Note that there is no assumption here that $E[q]$ is quadratic in $q$, though for many of the standard examples it is.

The key consequence is that the phase space measure is defined simply by replacing $\omega$ by $q$ in (\ref{3.4}). The statistical ensembles (\ref{3.18}) and (\ref{3.23}), and the form (\ref{3.30}) continues to define the Lagrange multiplier function $\mu(\sigma)$. The momentum term (\ref{2.12}) is also obtained by simply substituting $q$ for $\omega$. Formally this follows from the identity $\{q,L\} = \partial_\xi q$, where $\xi$ is the symmetry coordinate, which shows that $L$ is the generator of translations along $\xi$.

The variational result for the free energy proceeds by following the steps (\ref{4.2})--(\ref{4.10}), but with the replacement
\begin{eqnarray}
-E[\omega_0] &\to& E[q_0] - \int_{\cal D} d{\bf r} \psi_0({\bf r}) q_0({\bf r})
\nonumber \\
&\equiv& L[\psi_0]
\label{6.9}
\end{eqnarray}
in which $L[\psi_0]$ (given by the domain integral of $-\frac{1}{2} |\nabla \psi_0|^2$ for the Euler equation) is the Legendre transform of $E_[q_0]$, obtained by inverting the relation $\psi_0[q_0]$ [generalizing (\ref{2.6})] to obtain $q_0[\psi_0]$ [generalizing (\ref{4.4})] and substituting the result into the first line of (\ref{6.9}). This inverse relationship is also encoded in $L$ via the general Legendre transform relation
\begin{equation}
q_0({\bf r}) = -\frac{\delta L}{\delta \psi_0({\bf r})}.
\label{6.10}
\end{equation}
The free energy now generalizes to
\begin{equation}
{\cal F}[\psi_0;\beta,{\bm \mu}]
= L[\psi_0] + \int_{\cal D} d{\bf r}
W[\psi_0({\bf r}) - \mu_P \alpha({\bf r});\beta,{\bm \mu}]
\label{6.11}
\end{equation}
in which the Lagrange multiplier--Laplace transform $W(\tau)$ continues to be defined by (\ref{4.6}) and (\ref{4.9}). The local $q$-distribution function continues to take the from (\ref{4.11}), and the equilibrium equation (\ref{4.12}) generalizes to
\begin{equation}
q_0({\bf r}) = -\frac{\delta L}{\delta \psi_0({\bf r})}
= \int d\sigma n_0({\bf r},\sigma).
\label{6.12}
\end{equation}

\subsection{Quasi-geostrophic flow and nonlinear Rossby waves}
\label{sec:qgrossby}

Quasi-geostrophic (QG) flow, including the Coriolis term $f$ described in Sec.\ \ref{sec:rotbetaplane}, is defined by
\begin{eqnarray}
q({\bf r}) &=& \omega({\bf r}) + k_R^2 \psi({\bf r}) + f({\bf r})
\nonumber \\
&=& (-\nabla^2 + k_R^2) \psi({\bf r}) + f({\bf r})
\label{6.13}
\end{eqnarray}
in which $R_0 = 1/k_R = c/f$ is the Rossby radius of deformation, the length scale beyond which Coriolis effects begin to dominate gravitational/hydrostatic effects on the fluid dynamics, with $c$ the speed of internal gravity waves (typically a few m/s on Earth \cite{foot:rossby}). This model, whose large scale wave excitations are known as Rossby waves, emerges from the shallow water equations, discussed in Sec.\ \ref{sec:swwaveeddy}, in the limit where the surface height adiabatically follows the eddy motion via quasi-hydrostatic balance. Higher frequency traveling surface wave excitations are neglected. The energy function is
\begin{equation}
E[q] = \frac{1}{2} \int_{\cal D} d{\bf r} \int_{\cal D} d{\bf r}'
[q({\bf r}) - f({\bf r})] G_R({\bf r},{\bf r}') [q({\bf r}') - f({\bf r}')]
\label{6.14}
\end{equation}
in which the Green function now obeys the Poisson equation [compare (\ref{2.7})]
\begin{equation}
(-\nabla^2 + k_R^2) G_R({\bf r},{\bf r}') = \delta({\bf r}-{\bf r}').
\label{6.15}
\end{equation}
In free space one obtains the modified Bessel function form
\begin{equation}
G_R^0({\bf r},{\bf r}') = \frac{1}{2\pi} K_0(k_R|{\bf r}-{\bf r}'|),
\label{6.16}
\end{equation}
which maintains the Euler equation logarithmic singularity near the origin, but decays exponentially $\sim e^{-|{\bf r}-{\bf r}'|/R_0}$ on the scale of the Rossby radius (which depends strongly on latitude, but is on the order of 50 km at mid-latitudes on Earth). The free surface motions therefore act to screen the vortex charge at larger distances.  With this adjustment of $G$, the statistical functional (\ref{3.25}) continues to take the general form of the field theory displayed in the first row of Table \ref{tab:fields} (though $f$ has been dropped there for simplicity).

The Legendre transform operation yields the form
\begin{equation}
L[\psi] = \int_{\cal D} d{\bf r} \left[\frac{1}{2} |\nabla \psi|^2
+ \frac{1}{2} k_R^2 \psi^2 + f\psi \right].
\label{6.17}
\end{equation}
Again, the momentum functionals are identical to those of the Euler equation, with the same function $\alpha({\bf r})$ as appearing in (\ref{2.10})--(\ref{2.12}).

The equilibria of this system has been explored by a number of authors \cite{BS2002,W2006,BV2012}. An interesting aspect of the vortex screening, and resulting finite range interactions, is that eddies with size much larger than $R_0$ have identical physics as fluid droplets with finite surface tension. Thus, the transition between interior and exterior of the eddy occurs over length scale $R_0$, and a surface energy per unit length $\Sigma_0(R_0,\beta,{\bm \mu})$ (see, e.g., Fig.\ 1 in Ref.\ \cite{W2006}) may be assigned to this interface. The shape of the eddy is obtained by minimizing the total surface energy subject to the effective external forces provided by the Coriolis and angular momentum effects. In particular, the Coriolis term is analogous to an external gravitational field and the vorticity is analogous to a mass density. It follows that the equilibrium state will tend to organize with ``lighter'' regions of lower vorticity floating on (northwards of) ``heavier'' regions of higher vorticity. This provides a partial explanation for the ubiquity of ``zonal jet'' structures, with compact eddies requiring a rarer balance of forces.

\begin{figure}

\includegraphics[width=3.2in,viewport = 190 20 810 490,clip]{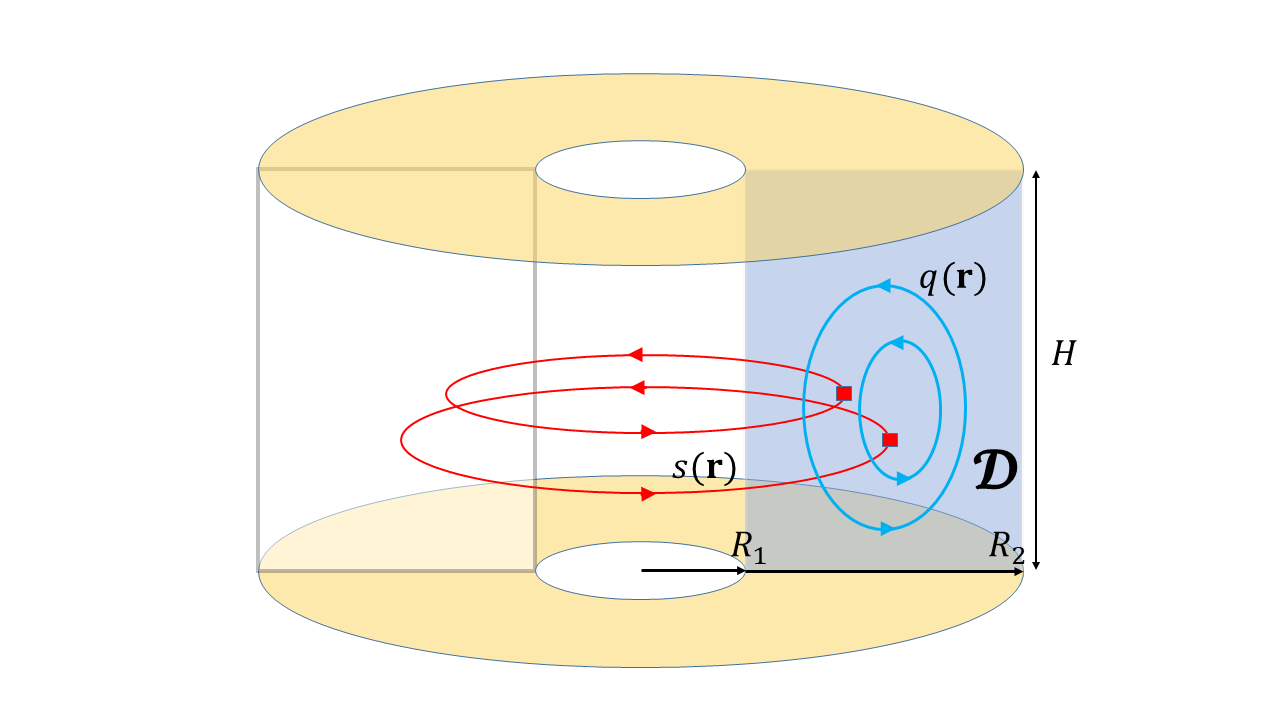}

\caption{Axisymmetric flow geometry confined to a cylinder of height $H$ and inner and outer radii $R_1 < R_2$. The pattern of flows is taken to be invariant under rotation about the cylinder axis, and is therefore specified by a toroidal flow field $s$ about the axis, and a poloidal vorticity field $q$ within any 2D radial planar section $D$.}

\label{fig:axisymcartoon}
\end{figure}

\subsection{Adiabatic conservation laws and slow equilibration}
\label{sec:adiabatconserve}

The QG equation has an added complication that, in addition to conservation laws treated so far, it has an additional approximate \emph{adiabatic} invariant $B[q]$ \cite{BNZ1991,BHW2011}. Like the energy, $B$ is quadratic in $q$, is insensitive to the microscale fluctuations, hence dominated by the large scale flow. Its conservation improves as the flow becomes more weakly nonlinear, hence very often as the turbulent state relaxes and the flow equilibrates. It is likely, therefore, that this invariant acts as a barrier to full equilibration---its approximately conserved value, computed from the initial state, will typically be different from that computed from the equilibrium state based on energy and Casimir conservation alone.

It may be argued, for example, that preserving $B$ constrains the inverse cascade to focus energy on wavevectors close to the $y$-axis (axis of rotation), hence (through the curl relation) enhancing the formation of zonal flows (organized normal to that axis). The equilibrium solutions (ignoring $B$) also often yield zonal flows, but the north--south geometry will in general be different. The full consequences of this competition deserve to be more fully explored.

\section{3D axisymmetric flow}
\label{sec:3daxisym}

Over the next few sections we will briefly review applications of equilibrium ideas to yet more complicated fluid systems. More details may be found in the referenced literature. The steps outlined in the previous sections---Liouville theorem and statistical measure, choice of equilibrium ensemble, entropy and free energy functions---remain highly relevant, but the exact variational solution derived for the Euler equation is in general no longer available. Rather, it becomes an approximate tool, along lines similar to the use of mean field theories in conventional systems. Specifically, the equilibrium states, though still constrained by an infinite number of conserved integrals, now contain further degrees of freedom (such as a free surface height or other additional coupled field) that escape the constraints, and continue to exhibit fluctuations on finite scales.

\subsection{Axisymmetric equation of motion}
\label{sec:axisymmotioneq}

The case of 3D axisymmetric flow, illustrated in Fig.\ \ref{fig:axisymcartoon}, will be our first example of the impact of an additional degree of freedom, not constrained by Casimirs \cite{TDB2014,W2019}. Under the constraint of azimuthal symmetry, and specializing to cylindrical coordinates, the full 3D Euler equation velocity field may be written in the form
\begin{eqnarray}
{\bf v} &=& v_r(r,z) \hat {\bf r} + v_z(r,z) \hat {\bf z}
+ v_\theta(r,z) \hat {\bm \theta}
\nonumber \\
&=& \nabla \times \left(\frac{1}{r} \psi \hat {\bm \theta} \right)
+ \frac{1}{r} s \hat {\bm \theta}
\label{7.1}
\end{eqnarray}
in which
\begin{equation}
s(r,z) = r v_\theta
\label{7.2}
\end{equation}
is the vertical component of the angular momentum density and characterizes ``toroidal'' flow around about the axis, while the 3D incompressibility condition allows one to express the ``poloidal'' flow components in terms of a stream function $\psi(r,z)$ via $v_r = -(\partial_z \psi)/r$, $v_z = (\partial_r \psi)/r$. The latter is related to the poloidal vorticity $\omega_\theta = \hat {\bm \theta} \cdot \nabla \times {\bf v}$ via
\begin{eqnarray}
q(r,z) \equiv \frac{\omega_\theta }{r}
&=& -\frac{1}{r^2} \partial_z^2 \psi
-\frac{1}{r} \partial_r \left(\frac{1}{r} \partial_r \psi \right)
\nonumber \\
&=& -\left(\frac{1}{2y} \partial_z^2 + \partial_y^2 \right) \psi
\equiv \Delta_* \psi
\label{7.3}
\end{eqnarray}
which serves to define a modified radial coordinate $y = r^2/2$ and modified 2D Laplacian $\Delta_*$. Defining the 2D coordinate ${\bm \rho} = (y,z)$, the formal inverse of the latter is obtained from the (Dirichlet) Green function relation
\begin{eqnarray}
\psi({\bm \rho}) &=& \int_{\cal D} d{\bm \rho}'
G({\bm \rho},{\bm \rho}') q({\bm \rho}')
\nonumber \\
-\Delta_* G({\bm \rho},{\bm \rho}') &=& \delta({\bm \rho}- {\bm \rho}').
\label{7.4}
\end{eqnarray}
generalizing (\ref{2.6}) and (\ref{2.7}).

Defining the modified 2D gradient $\nabla_\rho = (\partial_y, \partial_z)$ and velocity ${\bf w} = \nabla_\rho \times \psi = (rv_r, v_z)$, one obtains the incompressibility condition $\nabla_\rho \cdot {\bf w} = 0$, and the Euler equation may be reduced to the coupled pair of scalar equations
\begin{eqnarray}
\partial_t s + {\bf w} \cdot \nabla_\rho s &=& 0
\nonumber \\
\partial_t q + {\bf w} \cdot \nabla_\rho q &=& \frac{\partial_z s^2}{4y^2}.
\label{7.5}
\end{eqnarray}
The first states that the toroidal velocity field is in essence a passive scalar that is freely advected by the poloidal velocity field generated by $q$ and obtained from the curl of (\ref{7.4}). The second states that the self-advection of the poloidal vorticity field is additionally forced by $s$, a type of Coriolis effect. In the absence of such forcing the $q$ equation would be formally identical to the Euler equation (\ref{2.4}). The effects of this forcing play a critical role in the statistical equilibria, which therefore differ strongly from those of the Euler equation.

\subsection{Conservation laws}
\label{sec:axisymcons}

In addition to conservation of total (kinetic) energy
\begin{eqnarray}
E[q,s] &=& E_G[q] + E_0[s]
\nonumber \\
E_G[q] &=& \pi \int_{\cal D} d{\bm \rho} \int_{\cal D} d{\bm \rho}'
q({\bm \rho}) G({\bm \rho},{\bm \rho}') q({\bm \rho}')
\nonumber \\
E_0[s] &=& \pi \int_{\cal D} d{\bm \rho} \frac{s({\bm \rho})^2}{2y},
\label{7.6}
\end{eqnarray}
one obtains two classes of Casimir-type constraints. The first of (\ref{7.5}) leads directly to conservation of the domain integral any function of $F(s)$, which may be characterized by conservation of the function [compare (\ref{2.14})]
\begin{equation}
g(\sigma) = \int_{\cal D} d{\bm \rho} \delta[\sigma - s({\bm \rho})].
\label{7.7}
\end{equation}

For a strictly finite cylinder of height $H$, with Dirichlet boundary conditions on all surfaces, equation (\ref{7.7}) comprises all of the Casimirs---there is no constraint on $q$ at all. However, in the case of \emph{periodic boundary conditions} in $z$ (termed a Taylor--Couette type geometry) with specified period $H$, it follows the second of (\ref{7.5}) that the domain integral of any combination of the form $q f(s)$ is conserved as well, characterized by conservation of the function
\begin{equation}
\tilde g(\sigma) = \int_{\cal D} d{\bm \rho}
q({\bm \rho}) \delta[\sigma - s({\bm \rho})].
\label{7.8}
\end{equation}
In essence, Dirichlet boundary conditions impose additional forces on the vertical motion of the fluid that destroy this constraint. In most of what follows we will assume periodic boundary conditions since it leads to more interesting results.

The additional constraint (\ref{7.8}) implies that the \emph{mean value} of $q$ over each level set $s({\bf r}) = \sigma$ is conserved, but may otherwise fluctuate arbitrarily. In particular, there is no control over the range of values that $q$ may take, allowing for unbounded fluctuations about this mean.

\subsection{Axisymmetric equilibria}
\label{sec:axisymeq}

Along similar lines to that derived in Sec.\ \ref{subsec:liouvilleeuler}, the Liouville theorem leads to equilibrium measures that must take the form of a conserved integral, with phase space integral defined by the continuum limit of free integration over the $s$ and $q$ fields:
\begin{equation}
\int d\Gamma = \int D[q] \int D[s] = \lim_{a \to 0} \prod_i
\int_{-\infty}^\infty dq_i \int_{-\infty}^\infty ds_i.
\label{7.9}
\end{equation}
Details may be found in App.\ A of Ref.\ \cite{W2019}. The grand canonical ensemble is defined by
\begin{eqnarray}
\rho_\mathrm{GC}[q,s] &=& \frac{1}{Z_\mathrm{GC}} e^{-\beta_a {\cal K}[q,s]}
\nonumber \\
Z_\mathrm{GC} &=& \int D[q] \int D[s] e^{-\beta_a {\cal K}[q,s]}
\label{7.10}
\end{eqnarray}
with inverse temperature $\beta_a = \beta/a^2$ again scaling with $a$ [see (\ref{3.26})]. Lagrange multiplier functions $\mu(\sigma)$ and $\tilde \mu(\sigma)$ enforcing conservation of $g(\sigma)$ and $\tilde g(\sigma)$, respectively, are introduced through the grand canonical statistical functional
\begin{eqnarray}
{\cal K}[q,s] &=& E_G[q]
\label{7.11} \\
&&+\ \int_{\cal D} d{\bm \rho}
\left\{\frac{\pi}{2y} s({\bm \rho})^2 - \mu[s({\bm \rho})]
- q({\bm \rho}) \tilde \mu[s({\bm \rho})] \right\}.
\nonumber
\end{eqnarray}
This model, reproduced in the second row of Table \ref{tab:fields}, takes the form of a purely local field $s$, with no self interactions and $\mu(\sigma)$ playing the role of a local potential energy, linearly coupled to an unconstrained Gaussian field $q$. As such, its thermodynamic behavior bears little resemblance to that of the Euler equation (\ref{3.25}) with (\ref{2.9}) and (\ref{3.30}). We summarize here its basic properties---full details may again be found in Ref.\ \cite{W2019}.

The first observation is that the magnitude of $q$ is controlled only by the positive definite quadratic form $E_G[q]$. The linear term $q \tilde \mu(s)$ serves (by completing the square) only to shift the mean. This type of shift is exactly the degree of freedom required to enforce the second set of Casimirs (\ref{7.8}). Being quadratic, the resulting Gaussian statistical averages over $q$ are finite and well defined only for positive temperatures, $\beta > 0$. However, being Gaussian, arbitrarily high energy flows may be created at positive temperature, so all of the conservation laws continue to be satisfied. In contrast, for the 2D Euler equation negative temperatures may be required because the Casimir constraints also bound, through $\omega$, the maximum energy of positive temperature states.

The end result is that because fluctuations about the local mean in $q$ are uncontrolled, one obtains identically vanishing mean stream function $\psi_0({\bm \rho})$ and poloidal vorticity $q_0({\bm \rho}) = -\Delta_* \psi_0({\bm \rho})$. Hidden from these are the finite averages of higher order quantities, such as the mean square velocity $\langle |\nabla_\rho \psi({\bm \rho})|^2 \rangle \propto T > 0$ (an equipartition result). In this sense the equilibria are similar to those of conventional particle systems.

The second observation is that if the initial flow is such that $|s|$ is bounded, then $e^{\beta \mu(s)}$ will be as well. Thus statistical averages over the field $s$ are well defined irrespective of the value of $\beta$ (either positive and negative, though as seen $q$ requires $\beta > 0$). However the $s({\bm \rho})^2$ energy contribution from $s$ is purely local, and the long range Coulomb interaction effects seen in the 2D Euler case are absent here. Given the absence of any finite scale structure in $q$, the $q \tilde \mu(s)$ term may be shown to play a negligible role in the statistics of $s$, and one obtains the exact $q$-independent result
\begin{eqnarray}
p_s(\sigma,y) \equiv \langle \delta[\sigma - s({\bm \rho})] \rangle
&=& \frac{1}{Z_1(\beta,{\bm \mu},y)}
e^{-\beta [\pi \sigma^2/2y - \mu(\sigma)]}
\nonumber \\
Z_1(\beta,{\bm \mu},y) &=& \int d\sigma
e^{-\beta [\pi \sigma^2/2y - \mu(\sigma)]}
\label{7.12}
\end{eqnarray}
for the local distribution of $\sigma$. In particular the local mean is derived in the form
\begin{equation}
s_0(y) \equiv \langle s({\bm \rho}) \rangle
= \int d\sigma \sigma p_s(\sigma,y).
\label{7.13}
\end{equation}

%Include model figure for this result?

\subsection{Equilibration issues}
\label{sec:eqissue}

Just as for the Euler equation, there are significant questions regarding the convergence to equilibrium for Taylor--Couette flows of this type. In particular, experiments do appear to show very long lived negative temperature-type states, with $q$ displaying large scale coherent structure (see Refs.\ \cite{TDB2014,LDC2006,NMCD2010} and references therein). Reasonable comparisons with experiments were obtained in Ref.\ \cite{TDB2014} by artificially bounding $|q| < M$, with $M$ remaining finite in the continuum limit $a \to 0$, and applying Euler equation mean field ideas to obtain negative temperature states for the resulting altered model.

Elucidating the barriers to equilibration, limiting or greatly slowing the growth of $|q|$ predicted by the model (\ref{7.11}) as the forward cascade proceeds, remains an interesting open question. We will encounter very similar issues below in relation to the surface height field for the shallow water equations.

\begin{figure}

\includegraphics[width=3.2in,viewport = 260 240 660 370,clip]{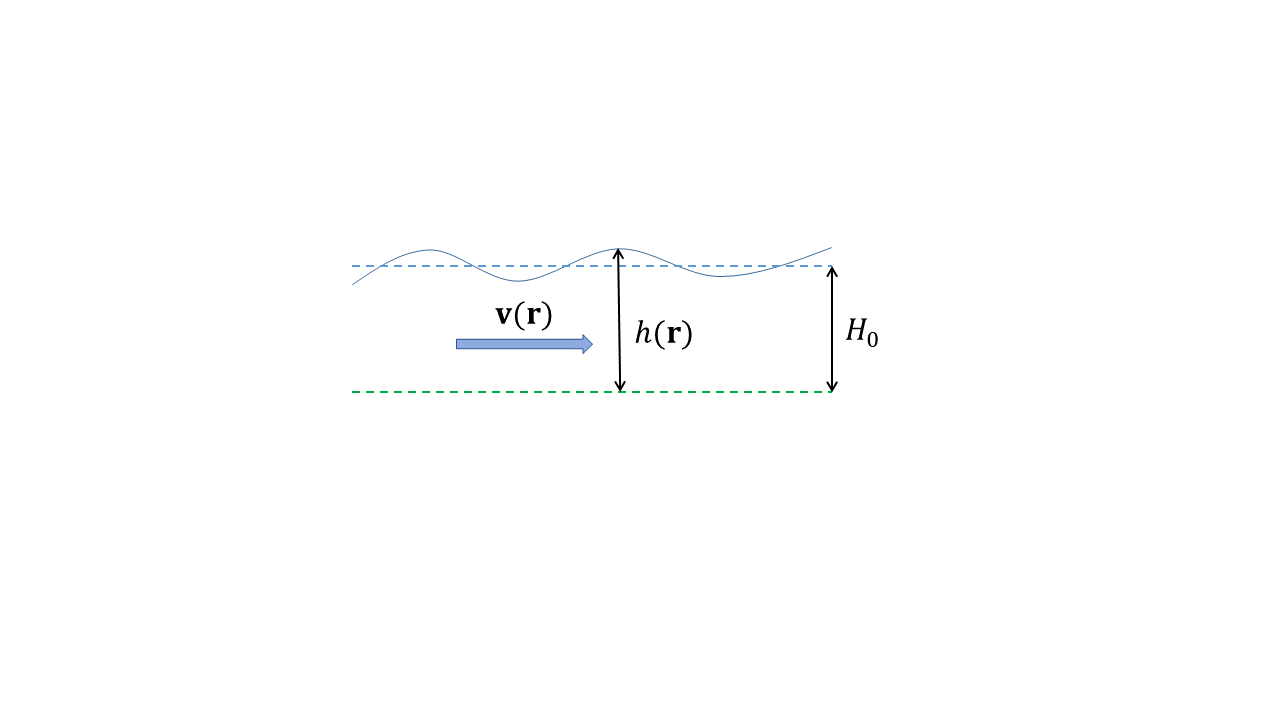}

\caption{Schematic illustration of the shallow water geometry.}

\label{fig:swcartoon}
\end{figure}

\section{Shallow water dynamics and wave-eddy interactions}
\label{sec:swwaveeddy}

The shallow water system, illustrated in Fig.\ \ref{fig:swcartoon}, is defined by the equations of motion
\begin{eqnarray}
\frac{D{\bf v}}{Dt} + {\bf v} \cdot \nabla {\bf v}
+ f \hat {\bf z} \times {\bf v} &=& -g \nabla h
\nonumber \\
\partial_t h + \nabla \cdot (h{\bf v}) &=& 0
\label{8.1}
\end{eqnarray}
in which ${\bf v}$ is the horizontal velocity, $h$ is the (fluctuating) fluid free surface height over a flat bottom, $z=0$, and for convenience we include the Coriolis parameter $f$ from the outset. Comparing to the Euler form (\ref{4.21}), the pressure gradient is provided by changes in surface height, and the second equation expresses incompressibility of the full 3D velocity by relating surface height change to the divergence of the mass current ${\bf j} = h{\bf v}$. These equations are derived from the 3D Euler equation in the formal asymptotic limit in which the length scale of horizontal variability (including the horizontal extent of the 2D domain ${\cal D}$) is much larger than $h$, and ${\bf v}$ is approximated as independent of $z$. The vertical velocity is then $v_z = -z \nabla \cdot {\bf v}$, hence $v_z({\bf r},h) = -h \nabla \cdot {\bf v}$, and it follows that the $h$ equation may be equivalently written in the intuitive form $Dh/Dt = v_z(h)$.

\subsection{Conservation laws}
\label{sec:swconserve}

It is straightforward to check that the potential vorticity
\begin{equation}
\Omega({\bf r}) = \frac{\omega({\bf r}) + f({\bf r})}{h({\bf r})}
\label{8.2}
\end{equation}
is advectively conserved, $D\Omega/Dt = 0$, which clearly reduces to (\ref{4.22}) for fixed surface height. The corresponding conserved Casimir area integrals are
\begin{equation}
g(\sigma) = \int_{\cal D} d{\bf r}
h({\bf r}) \delta[\sigma - \Omega({\bf r})].
\label{8.3}
\end{equation}
Integrating this over $\sigma$, one obtains in particular conservation of the mean height
\begin{equation}
H_0 = \frac{1}{A_{\cal D}} \int d\sigma g(\sigma)
= \frac{1}{A_{\cal D}} \int_{\cal D} d{\bf r} h({\bf r}).
\label{8.4}
\end{equation}

Since the fluid is compressible, the additional compression field
\begin{equation}
Q({\bf r}) = \frac{q({\bf r})}{h({\bf r})},\ \
q({\bf r}) \equiv \nabla \cdot {\bf v}({\bf r})
\label{8.5}
\end{equation}
is required to fully reconstruct the velocity in the form
\begin{equation}
{\bf v} = \nabla \times \psi - \nabla \phi
\label{8.6}
\end{equation}
in which the stream function $\psi$ and potential function $\phi$ are obtained by solving
\begin{equation}
\omega = h\Omega + f = -\nabla^2 \psi,\ \ q = hQ = -\nabla^2 \phi.
\label{8.7}
\end{equation}
Free slip boundary conditions on ${\bf v}$ require as before Dirichlet boundary conditions on the stream function $\psi$, but Neumann boundary conditions $\hat {\bf n} \cdot \nabla \phi|_{\partial {\cal D}} = 0$, on the compression potential. Thus, one obtains
\begin{eqnarray}
\psi({\bf r}) &=& \int_{\cal D} d{\bf r}' G_D({\bf r},{\bf r}') \omega({\bf r}')
\nonumber \\
\phi({\bf r}) &=& \int_{\cal D} d{\bf r}' G_N({\bf r},{\bf r}') q({\bf r}')
\label{8.8}
\end{eqnarray}
with subscripts labeling the Green function boundary conditions. Both are long-ranged, with logarithmic singularities at the origin.

The conserved energy
\begin{eqnarray}
E &=& E_K + E_P
\nonumber \\
E_K &=& \frac{1}{2} \int_{\cal D} d{\bf r}
h({\bf r}) |{\bf v}({\bf r})|^2
\nonumber \\
E_P &=& \frac{1}{2} g \int_{\cal D} d{\bf r} h({\bf r})^2
\label{8.9}
\end{eqnarray}
is a sum of kinetic and potential terms. By substituting (\ref{8.6}) and (\ref{8.8}), the kinetic term may be organized in the form
\begin{eqnarray}
E_K[\Omega,Q,h] &=& \frac{1}{2} \int_{\cal D} d{\bf r} \int_{\cal D} d{\bf r}'
\left[\begin{array}{c} (h\Omega-f)({\bf r}) \\ hQ({\bf r}) \end{array} \right]^T
\nonumber \\
&&\times\ {\cal G}_h({\bf r},{\bf r}')
\left[\begin{array}{c} (h\Omega-f)({\bf r}') \\ hQ({\bf r}') \end{array} \right]
\label{8.10}
\end{eqnarray}
in which the components of the $2\times 2$ tensor Green function ${\cal G}_h({\bf r},{\bf r}')$ is an integral-product of $h$ with gradients of $G_D$ and $G_N$. The exact from is not needed in what follows since expressions in terms of $\psi$ will reemerge as central in the statistical analysis. In the presence of translation or rotation symmetry, momentum conservation analogous to (\ref{2.10}) or (\ref{2.11}) also occurs, but will be neglected here for simplicity.

\subsection{Liouville theorem and statistical measures}
\label{sec:swliouville}

Proving a Liouville theorem for this system is much more involved, and we only quote the result here---full derivation may be found in App.\ A of Ref.\ \cite{W2017}. The simplest approach, conceptually, is to treat the height $h$ and the two components of the mass current ${\bf j} = h{\bf v}$ as fundamental canonical variables. In terms of these it can be shown that the correct phase space integration measure, accompanying the conserved equilibrium density $\rho_\mathrm{eq}[h,{\bf j}]$, continues to be defined by approximating the domain ${\cal D}$ by a uniform mesh, with lattice parameter $a$, and freely integrating over the discretized fields:
\begin{eqnarray}
\int d\Gamma &=& \int D[h] \int D[{\bf j}]
\equiv \lim_{a \to 0} \prod_i \int dh_i \int d{\bf j}_i
\nonumber \\
&=& \lim_{a \to 0} \prod_i \int h_i^2 dh_i \int d{\bf v}_i.
\label{8.11}
\end{eqnarray}
From this representation, using finite difference approximations to the gradients, one may change variables from ${\bf v}$ to $(\Omega,Q)$ to obtain
\begin{eqnarray}
\int d\Gamma &=& \lim_{a \to 0} \prod_i a^2
\int h_i^2 dh_i \int dq_i \int d\omega_i
\nonumber \\
&=& \lim_{a \to 0} \prod_i a^2
\int h_i^4 dh_i \int dQ_i \int d\Omega_i
\nonumber \\
&\equiv& \int D[h] \int D[\Omega] \int D[Q]
\label{8.12}
\end{eqnarray}
The only slightly usual feature is the height measure $h^4 dh$ coming from the various changes of variable.

\subsection{Shallow water equilibria}
\label{swequilibria}

The grand canonical form of the equilibrium measure $\rho_\mathrm{eq} = Z^{-1} e^{-\beta_a {\cal K}}$ is again obtained by introducing the Lagrange multiplier function $\mu(\sigma)$ to control the Casimir constraints. The statistical functional is
\begin{equation}
{\cal K}[\Omega,Q,h] = E_K[\Omega,Q,h]
+ \int_{\cal D} \left\{\frac{1}{2} g h({\bf r})^2
- h({\bf r}) \mu[\Omega({\bf r})] \right\},
\label{8.13}
\end{equation}
also displayed, with a more compact notation, in the third row of Table \ref{tab:fields}. Like the Euler equation, and unlike for the axisymmetric flow model, the vortex degree of freedom $\Omega$ is directly constrained by the Casimirs. However, the additional (height and compression) degrees of freedom enter in a complicated way that makes this model very difficult to analyze. Height fluctuations, controlled only locally by the $\frac{1}{2} gh^2$ term, are strongly coupled to $\Omega$, and forbid any simple reduction to a mean field type description.

In order to gain some intuition and make closest possible contact with the Euler equation, one may integrate out the Gaussian field $Q$ to obtain the reduced functional
\begin{eqnarray}
\hat {\cal K}[\Omega,h] &=& E_K[\Omega,h]
+ \int d{\bf r} \left\{\frac{1}{2} g h({\bf r})^2
- h({\bf r}) \mu[\Omega({\bf r})] \right\}
\nonumber \\
\hat E_K[\Omega,h] &\equiv& \frac{1}{2} \int_{\cal D} d{\bf r}
\int_{\cal D} d{\bf r}' (h\Omega-f)({\bf r})
\nonumber \\
&&\times\ G_h({\bf r},{\bf r}') (h\Omega-f)({\bf r}')
\label{8.14}
\end{eqnarray}
in which the scalar Green function satisfies
\begin{equation}
-\left(\nabla \cdot \frac{1}{h({\bf r})} \nabla \right)
G_h({\bf r},{\bf r}') = \delta({\bf r}-{\bf r}'),
\label{8.15}
\end{equation}
with Dirichlet boundary conditions. The resemblance to (\ref{3.25}), together with (\ref{2.9}) and (\ref{3.30}), is clear. However, the presence of the rapidly varying, not necessarily low amplitude, height field, without any intrinsic correlations that might perhaps smooth it out, drastically effects $G_h$. In particular, it is not smooth and hence strongly violates the conditions under which the mean field approximation described in Sec.\ \ref{sec:mfapproach} is derived. One may think of $G_h$ as generating a Coulomb-type interaction between vortices that retains a strong equilibrium fluctuation on finite length scales. Moreover, the $\frac{1}{2} gh^2$ term makes sense only at positive temperatures. Similar to the $q$ field in the 3D axisymmetric model, height fluctuations absorb unbounded energy for increasing $T$, and are hence in principle capable of dissipating negative temperature-like vortex states and converting large scale vortex motion into height fluctuations.

\subsection{Quasi-hydrostatic shallow water equilibria}
\label{sec:quasihydrosweq}

There are, however, physical motivations, completely outside of equilibrium considerations, for seeking equilibria with smooth height fields. Thus, a forward-type cascade of high amplitude, small scale height fluctuations will eventually violate the long wavelength assumption entering the derivation of the shallow water equations. When these assumptions are violated the full 3D Euler equations will display shock wave formation, wave breaking, and other 3D motions that will serve to effectively dissipate strong wave motions without significantly impacting large scale eddy motions.

Interesting work for the future would be more careful investigations of the validity of such alternative routes to equilibrium. For now let us briefly explore the consequences. If the height is smooth on the scale of variation of $\Omega({\bf r})$, then $G_h$ is smooth and, following steps analogous to the functional Taylor expansion (\ref{4.2}), one obtains the mean field approximation
\begin{eqnarray}
\hat {\cal K}[\Omega,h] &=& - \hat E_K[\Omega_0,h]
+ \int_{\cal D} d{\bf r} \left[\frac{1}{2} g h({\bf r})^2
- \Psi_0({\bf r}) f({\bf r}) \right]
\nonumber \\
&&+\ \int_{\cal D} d{\bf r} h({\bf r}) \left[\Psi_0({\bf r}) \Omega({\bf r})
- \mu[\Omega({\bf r})] \right],
\label{8.16}
\end{eqnarray}
in which the shallow water stream function $\Psi$ associated with the mass current $h{\bf v}$ (which is indeed incompressible in equilibrium, and differs from the velocity stream function $\psi$ introduced earlier), is defined by
\begin{equation}
-\left(\nabla \cdot\frac{1}{h({\bf r})} \nabla \right) \Psi_0({\bf r})
= (h\Omega_0 - f)({\bf r}) = \omega_0({\bf r}),
\label{8.17}
\end{equation}
leading to
\begin{equation}
\Psi_0({\bf r}) = \int_{\cal D} d{\bf r}' G_h({\bf r},{\bf r}')
(h\Omega_0 - f)({\bf r}').
\label{8.18}
\end{equation}
Using (\ref{8.15}) and (\ref{8.18}) one may express
\begin{eqnarray}
\hat E_K[\Omega_0,h] &=& \frac{1}{2} \int_{\cal D} d{\bf r}
\Psi_0({\bf r}) (h\Omega -f)({\bf r})
\nonumber \\
&=& -\frac{1}{2} \int_{\cal D} d{\bf r}
\Psi_0({\bf r}) \left(\nabla \cdot \frac{1}{h({\bf r})} \nabla \right) \Psi_0({\bf r})
\nonumber \\
&=& \int_{\cal D} d{\bf r} \frac{|\nabla \Psi_0({\bf r})|^2}{2h({\bf r})}
\label{8.19}
\end{eqnarray}

The fully fluctuating field $\Omega$ now appears only in the final local term in (\ref{8.16}), and one may now integrate it out to obtain the shallow water Free energy functional generalizing (\ref{4.10}):
\begin{eqnarray}
{\cal F}[\Psi_0,h;\beta,{\bm \mu}] &=& -E_K[\Omega_0,h]
+ \int_{\cal D} d{\bf r} W[h({\bf r}),\Psi_0({\bf r})]
\nonumber \\
&=& \int_{\cal D} d{\bf r} \left\{W[h({\bf r}),\Psi_0({\bf r})]
- \frac{|\nabla \Psi_0({\bf r})|^2}{2h({\bf r})} \right.
\nonumber \\
&&\left.+ \frac{1}{2} g h({\bf r})^2 - f({\bf r}) \Psi_0({\bf r}) \right\}
\label{8.20}
\end{eqnarray}
in which, generalizing (\ref{4.8}) and (\ref{4.9}), we define
\begin{equation}
e^{-\beta W(\tau,h)} = h^2 \int d\sigma e^{-\beta h [\tau \sigma - \mu(\sigma)]}
\label{8.21}
\end{equation}
where the $h^2$ prefactor comes from the phase space measure [the original $h_i^4$ in (\ref{8.12}) is reduced to $h_i^2$ after performing the $Q$ integral]. We again observe the required scaling $\beta_a = \beta/a^2$ to obtain a finite result in the continuum limit.

The self-consistent equation for $\Psi_0({\bf r})$ is obtained from the extremum condition $\delta {\cal F}/\delta \Psi_0({\bf r}) = 0$, which yields
\begin{eqnarray}
\omega_0({\bf r}) &\equiv& \langle \omega({\bf r}) \rangle
= -\left(\nabla \cdot\frac{1}{h({\bf r})} \nabla \right)
\Psi_0({\bf r}) + f({\bf r})
\nonumber \\
&=& \partial_\tau W[h({\bf r}),\Psi_0({\bf r})].
\label{8.22}
\end{eqnarray}
Similar to Euler result (\ref{4.12}), the self-consistency condition equates the mean vorticity derived from the equilibrium stream function [first line of (\ref{8.22})] with that computed from the local distribution function [second line of (\ref{8.22})], here emerging as a certain function $W$ of $\Psi_0$ controlled by the Lagrange multipliers $\beta,\mu(\sigma)$.

The equation for $h$ is obtained by applying the extremum condition $\delta {\cal F}/\delta h({\bf r}) = 0$,
\begin{eqnarray}
\frac{|\nabla \Psi_0({\bf r})|^2}{2h({\bf r})^2} + g h({\bf r})
&\equiv& \frac{1}{2} |{\bf v}_0({\bf r})|^2 + gh({\bf r})
\nonumber \\
&=& -\partial_h W[h({\bf r}),\Psi_0({\bf r})].
\label{8.23}
\end{eqnarray}
which corresponds to the reasonable assumption that the dissipation process self consistently acts to minimize the free energy. This is formally correct for large $\beta$ where height fluctuations are indeed small. Thus, more formally, the self-consistency requirement is that the dissipation process produces a new effectively low temperature system. The Lagrange multipliers $\mu(\sigma)$ will change as well so as to enforce the approximately the same $g(\sigma)$---to the extent the large scale eddy degrees of freedom are unaffected by the high frequency wave suppression.

In the large $\beta$ limit one can show that $\partial_h W(h,\Psi_0) \simeq W_1(\Psi_0)$ is independent of $h$. It follows then that the (\ref{8.23}) expresses the Bernoulli condition, namely that the sum of local kinetic energy and pressure is constant along stream lines (level curves of $\Psi_0$). This is indeed a rigorous requirement for steady flows. More generally, one may continue to apply (\ref{8.23}) for moderate values of $\beta$ as an approximate model in which some fluctuations in $h$ are kept (and the Bernoulli condition is weakly violated).

Another interesting consequence is that, accepting (\ref{8.20}) and an approximate free energy, negative temperatures are no longer precluded. Thus, $W(h,\tau)$ is perfectly well defined for $\beta < 0$ and solutions to (\ref{8.23}) may be sought for both positive and negative $\beta$. As previously stated, negative temperature equilibria are formally unstable to leakage of energy into (positive-temperature) wave motions, but the physical coupling of large-scale flows to small-scale wave generation is extremely weak and it makes sense to develop a theory along these lines that neglects such effects. The key observation here is that compact eddy structures, such as Jupiter's Great Red Spot, having vorticity maxima confined away from the system boundaries, can only be interpreted as negative-temperature states. Such structures therefore lie outside the strict shallow water theory presented here and nonequilibrium dissipation arguments must therefore be invoked in order to make contact with the effective equilibrium descriptions ubiquitous in the literature \cite{BV2012}.

The result (\ref{8.20}) reduces to the Euler equation result (\ref{4.10}) if one constrains $h({\bf r}) = H_0$. As one relaxes this constraint the vorticity pattern will evolve somewhat to accommodate the sloping surface in response to quasi-hydrostatic force balance, as observed in \cite{WP2001,CS2002}. However, one does not expect major changes from the 2D Euler result unless one drives the system to extremely high vorticity gradients, which is typically not of geophysical relevance.

\begin{figure}

\includegraphics[width=3.2in]{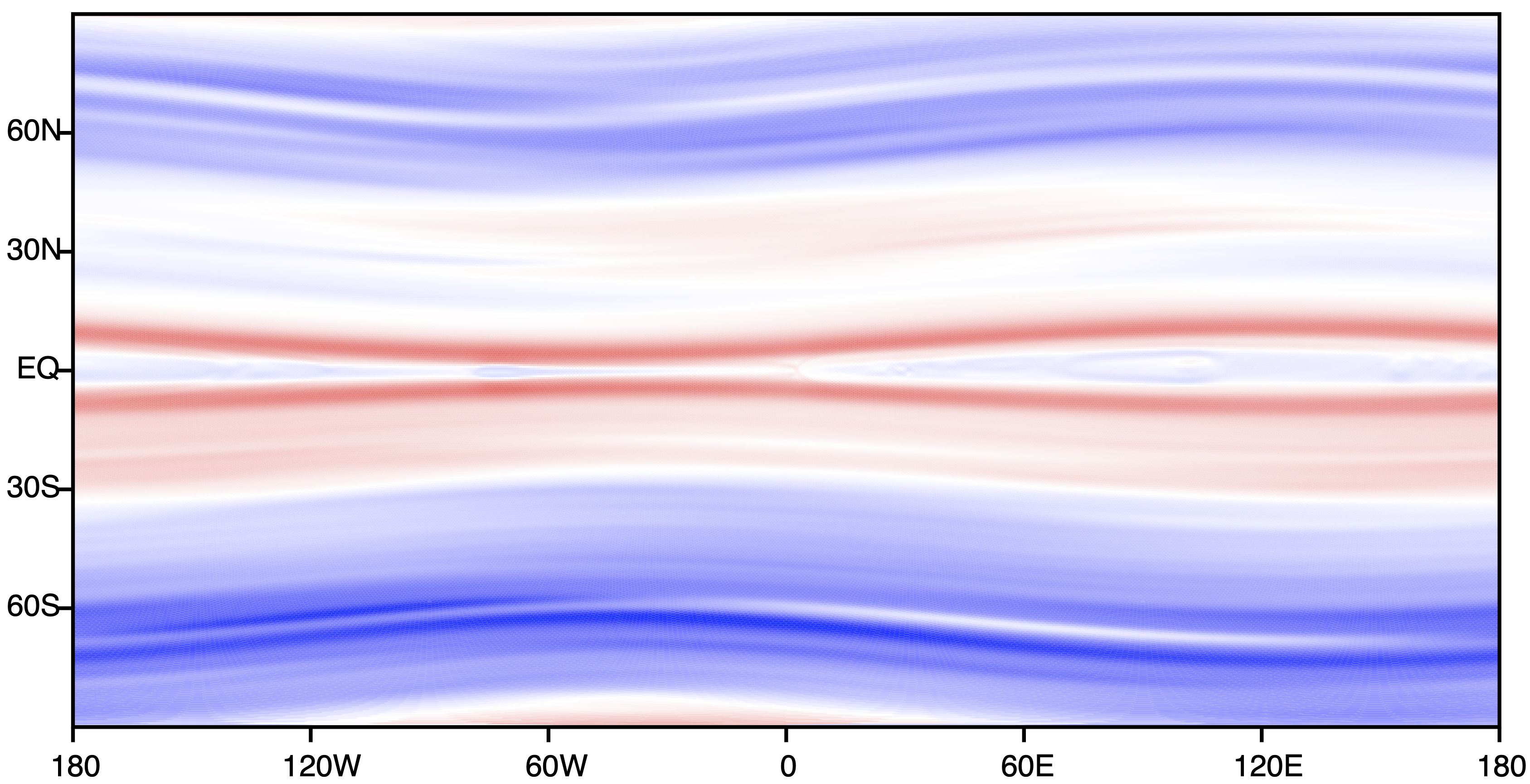}

\includegraphics[width=3.2in]{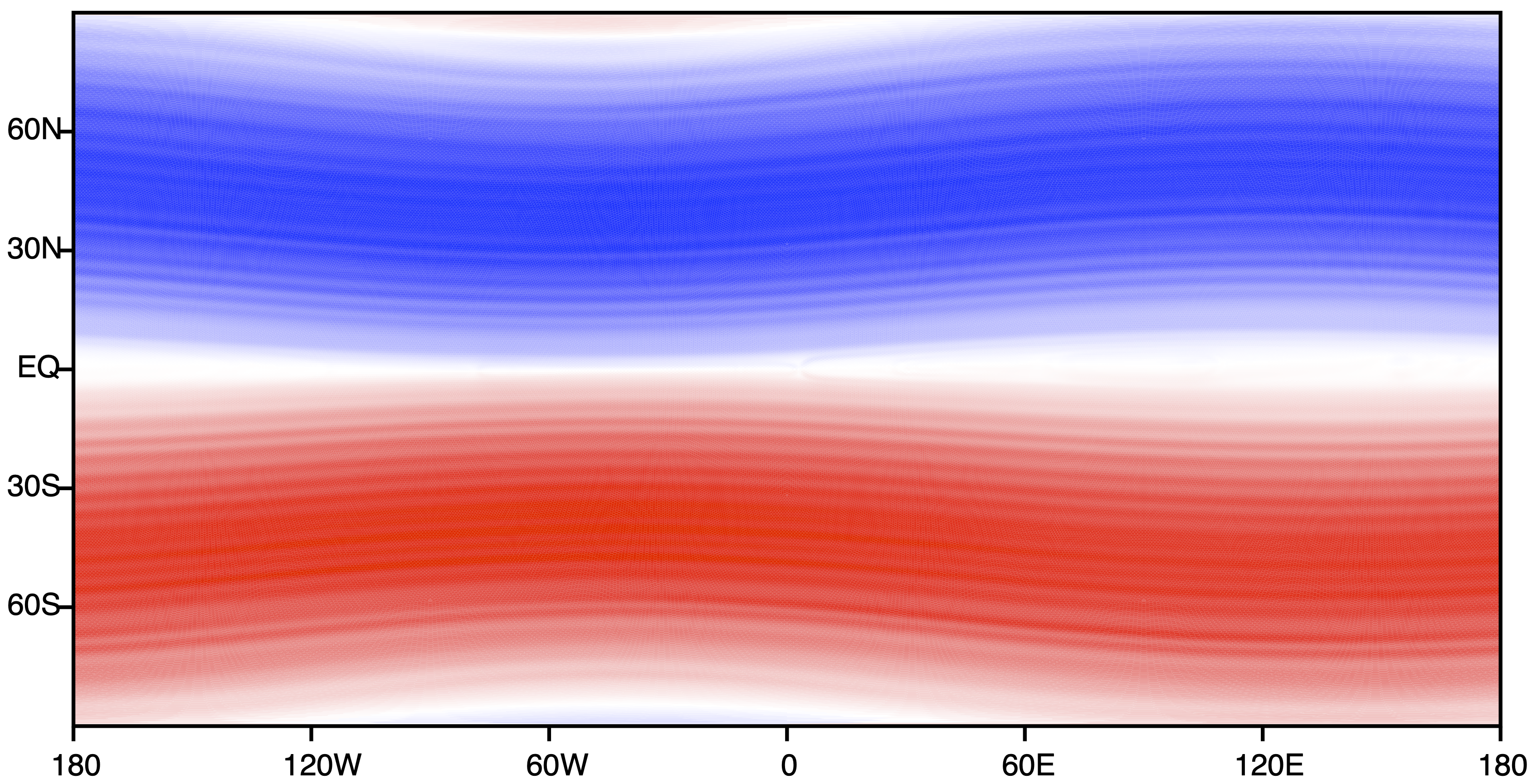}

\caption{Example numerically generated long-time (near-equilibrium) behavior of freely decaying 2D magnetohydrodynamics on the sphere. The zonal velocity field (above) and zonal magnetic field (below) undergo coupled dynamics according to (\ref{9.1}), reducing to (\ref{9.2}) and (\ref{9.3}) in the 2D solar tachocline model \cite{HMRW1985,TDH2007,PMT2019}. In the $(\psi,A)$ representation (\ref{9.8}) of the statistical functional, where ${\bf v}$ is defined by the level curves of $\psi$ and ${\bf B}$ is defined by the level curves of $A$, the model is that of two gradient-coupled membranes in an external potential which, among other things, tends to preferentially align the two vector fields.}

\label{fig:tachocline}

\end{figure}

\section{2D magnetohydrodynamics and the solar tachocline}
\label{sec:2Dmhd}

Our final example is that of the ideal, perfectly electrically conducting fluid (relevant to the energy conserving limit) interacting with an external magnetic field. The effective 2D theory of interest here emerges as follows. We begin with the 3D magnetohydrodynamic (MHD) equations
\begin{eqnarray}
\frac{D{\bf v}}{Dt} + {\bm \Omega} \times {\bf v}
&=& -\nabla P + {\bf J} \times {\bf B}
\nonumber \\
\partial_t {\bf B} &=& \nabla \times ({\bf v} \times {\bf B})
\label{9.1}
\end{eqnarray}
where ${\bm \Omega}$ is the rotation vector. The right hand side of the first equation now includes, in addition to the pressure term, the Lorentz force acting on a parcel of fluid. The second equation is Faraday's law with electric field determined by the constraint ${\bf E} + {\bf v} \times {\bf B} = 0$, which zeroes out the net force on the charge imposed by the perfectly conducting limit. The equations are closed using Ampere's law ${\bf J} = \nabla \times {\bf B}$, and the generalized pressure $P$ (which includes also contributions from centrifugal force, gravity, etc.)\ continues to enforce the incompressibility condition (\ref{2.2}). The constraint $\nabla \cdot {\bf B} = 0$ is automatically enforced by the second equation.

The solar tachocline (an example simulation result for which is shown in Fig.\ \ref{fig:tachocline}) is the observed sharp radial boundary between the solid body rotating radiative interior and differentially rotating outer convective zone. Here the current ${\bf J} = J \hat {\bf z}$ passes normally through the surface, while ${\bf v}$ and ${\bf B}$ are in-plane. The incompressibility conditions allow one to define the stream function (\ref{2.3}) together with the ($z$ component of the) magnetic vector potential
\begin{equation}
{\bf B} = \nabla \times A,\ \ J = -\nabla^2 A.
\label{9.2}
\end{equation}
The vector equations (\ref{9.1}) now reduce to the pair of scalar equations
\begin{equation}
\frac{D(\omega + f)}{Dt} = {\bf B} \cdot \nabla J,\ \
\frac{DA}{Dt} = 0.
\label{9.3}
\end{equation}
with $f = 2\Omega \sin(\varphi)$ defined by the solar latitude $\varphi$. The kinetic plus electromagnetic energy
\begin{eqnarray}
E &=& \frac{1}{2} \int_{\cal D} d{\bf r} (|{\bf v}|^2 + |{\bf B}|^2)
\nonumber \\
&=& \frac{1}{2} \int_{\cal D} d{\bf r} (|\nabla \psi|^2 + |\nabla A|^2)
\label{9.4}
\end{eqnarray}
is conserved if both $\psi$ and $A$ obey Dirichlet (free slip) boundary conditions. For annular or periodic strip geometries the angular momentum (\ref{2.12}) is conserved, and can alternatively be written in the form
\begin{equation}
P = - \int_{\cal D} d{\bf r} \nabla \alpha \cdot \nabla \psi.
\label{9.5}
\end{equation}

It is immediately evident that the Casimirs are completely different here since the potential vorticity $\omega + f$ is no longer advectively conserved. Instead it is the potential $A$ that is conserved, which has the drastic effect of imposing no direct control on the second derivative $J$. In fact, similar to the axisymmetric case (\ref{7.8}), there are two sets of Casimirs
\begin{equation}
g(\sigma) = \int_{\cal D} \delta[\sigma - A({\bf r})],\ \
\tilde g(\sigma) = \int_{\cal D} (\omega + f) \delta[\sigma - A({\bf r})]
\label{9.6}
\end{equation}
with the second following from the fact that $J{\bf B}$ is orthogonal to $\nabla A$. Dynamically, if ${\bf B} \cdot \nabla J$ happens to be small, one may expect to observe gradual evolution from Euler-type large scale eddy states to the quite different equilibria based on the vector potential Casimirs. The latter in particular permit diverging small scale vorticity fluctuations, as exhibited below. The presence of even weak magnetic field in 2D MHD simulations has indeed been found to destroy the conventional inverse cascade, breaking up large scale eddy flows \cite{TDH2007}. However, as described below, the new set of Casimirs (\ref{9.6}) are also capable of generating large scale flows, but based on significantly different initial states with imposed structure on $A$ rather than on $\omega$.

A Liouville theorem may straightforwardly be proven for the pair $\omega,A$ so that the equilibrium phase space measure is
\begin{eqnarray}
\int D[\omega] \int D[A] &=& \lim_{a\to 0} \prod_i \int d\omega_i \int dA_i
\nonumber \\
&=& \lim_{a\to 0} {\cal J}_a \prod_i \int d\psi_i \int dA_i
\label{9.7}
\end{eqnarray}
in which ${\cal J}_a$ is the Jacobian associated with the change of variable $\omega \to \psi$. This simply adds a constant to the free energy and drops out of any statistical average.

Defining corresponding Lagrange multipliers $\mu(\sigma)$ and $\tilde \mu(\sigma)$, we consider then the grand canonical statistical functional
\begin{eqnarray}
{\cal K}[A,\psi] &=& \int_{\cal D} d{\bf r} \bigg\{\frac{1}{2}|\nabla A|^2
+ \frac{1}{2}|\nabla \psi|^2 + \lambda \nabla \alpha \cdot \nabla \psi
\nonumber \\
&&-\ \tilde \mu'(A) \nabla A \cdot \nabla \psi
- [\mu(A) + f \tilde \mu(A)] \bigg\} \ \ \ \ \ \
\label{9.8}
\end{eqnarray}
in which integration by parts has been used to express everything in terms of at most first order gradients of the fields. This form is also displayed in the fourth row of Table \ref{tab:fields} (with $f$ again dropped for simplicity).

The physical model associated with $\rho_\mathrm{eq} = Z_\mathrm{GC}^{-1} e^{-\beta_a {\cal K}}$ is that of two membranes with ``heights'' $A({\bf r}), \psi({\bf r})$ and unit surface tension (the coefficient of the gradient-squared terms), and additionally coupled through their gradients. The term $\mu(A) + f({\bf r}) \tilde \mu(A)$ is a smoothly position-dependent external potential, confining $A$ near its minimum. The $\psi$ membrane is not directly confined, but the gradient coupling favors ${\bf B}$ parallel to $\tilde \mu'(A) {\bf v} + \lambda \nabla \times \alpha$.

Using the scaling $\beta_a = \beta/a^2$ one sees that the membrane experiences local Brownian-like fluctuations, with neighboring height differences scaling as $a/\sqrt{\beta}$. It follows that $A$ and $\psi$ are continuous, but have randomly fluctuating gradient, so that ${\bf v}$ and ${\bf B}$ fluctuate from site to site with scale $1/\sqrt{\beta}$. The membranes are therefore globally smooth but microscopically rough. In fact one may make use of this separation of scales to write
\begin{equation}
A = A_0 + \delta A,\ \ \psi = \psi_0 + \delta \psi
\label{9.9}
\end{equation}
in which $A_0,\psi_0$ are the equilibrium averages, to be determined self-consistently below, and $\delta A,\delta \psi = O(a/\sqrt{\beta})$ are fluctuation corrections. Substituting these into (\ref{9.8}) one obtains
\begin{eqnarray}
{\cal K}[A,\psi] &=& {\cal K}[A_0,\psi_0]
+ {\cal K}_2[\delta A,\delta \psi; A_0] + O(a)
\nonumber \\
{\cal K}_2[\delta A,\delta \psi; A_0]
&=& \frac{1}{2} \int_{\cal D} d{\bf r}
[|\nabla \delta\psi|^2 + |\nabla \delta A|^2
\nonumber \\
&&-\ 2\nu_0'(A_0) \nabla \delta A \cdot \nabla \delta\psi]
\nonumber \\
&=& \frac{1}{2} \int_{\cal D} d{\bf r}
\{[1 + \nu'(A_0)] |\delta \phi^-|^2
\nonumber \\
&&+\ [1 - \nu'(A_0)] |\delta \phi^+|^2] \}
\label{9.10}
\end{eqnarray}
in which $\delta \phi^\pm = (\delta A \pm \delta \psi)/\sqrt{2}$ are independent Gaussian fields. For smooth $A_0, \psi_0$, all other terms, including those linear in $\delta A,\delta \psi$, vanish with $a \to 0$. The major complication here is that the coefficient $\nu'[A_0({\bf r})]$ is not only position dependent, but yet to be determined.

The free energy functional follows from (\ref{9.9}) in the form
\begin{equation}
{\cal F}[A_0,\psi_0] = {\cal K}[A_0,\psi_0] + {\cal F}_2[A_0]
\label{9.11}
\end{equation}
in which the Gaussian correction is defined by
\begin{equation}
e^{-\beta_a {\cal F}_2[A_0]}
= \int D[\delta A] \int D[\delta \psi]
e^{-\beta_a {\cal K}_2[\delta A,\delta \psi;A_0]}
\label{9.12}
\end{equation}
and has a well defined continuum limit. The equilibrium equations, obtained from $\delta {\cal K}_0/\delta \psi_0({\bf r}) = 0$, $\delta {\cal F}/\delta A_0({\bf r}) = 0$ yield, respectively
\begin{eqnarray}
{\bf v}_0 &=& \tilde\mu'(A_0) {\bf B}_0 - \lambda \nabla \times \alpha
\label{9.13} \\
\omega_0 + \nu'(A_0)J_0 &=& \mu'(A_0) + f\tilde \mu'(A_0)
+ \tilde \mu''(A_0) \gamma({\bf r},{\bf r};A_0)
\nonumber
\end{eqnarray}
in which
\begin{equation}
\gamma({\bf r},{\bf r}') = \langle \delta A({\bf r})
\cdot \delta \psi({\bf r}') \rangle_2
= \langle \delta {\bf B}({\bf r}) \cdot \delta {\bf v}({\bf r}') \rangle_2
\label{9.14}
\end{equation}
is the magnetic--velocity Gaussian correlation function. The first equation provides a direct relation between the equilibrium velocity and magnetic field, being collinear up to a momentum conservation-induced mean flow subtraction---this is the effect of the gradient coupling term in (\ref{9.8}). By substituting the curl of this relation into the second equation, it is straightforward to derive a closed equation for $A_0$ alone.

These equations look quite complicated, but have a straightforward physical interpretation. The functional ${\cal K}[A_0,\psi_0]$ reflects a classical surface tension minimization problem in the presence of the external potentials $\mu,\tilde \mu$. The second derivative term $\omega_0 + \nu'(A_0) J_0$ in the second equation represents a surface tension restoring force in response to the forcing terms on the right. The $\gamma$ term represents the fluctuation corrections to the surface tension due to the membrane roughening effect. The self-consistent dependence on $A_0,\psi_0$ arises from such effects as regions of strongly stretched membrane having reduced amplitude fluctuations. Example solutions of these equations, displaying similar large scale vortex flow patterns as the Euler equations, are shown in Ref.\ \cite{W2012}.

All equilibrium conserved integrals are derived through differentiation with respect to the Lagrange multipliers as usual:
\begin{eqnarray}
E &=& \frac{1}{2} \int_{\cal D} d{\bf r}
[|{\bf v}_0|^2 + |{\bf B}_0|^2 + \varepsilon({\bf r})]
\nonumber \\
L &=& -\frac{\partial {\cal F}}{\partial \lambda}
= \int_{\cal D} d{\bf r} \alpha \omega_0
\nonumber \\
g(\sigma) &=& -\frac{\delta {\cal F}}{\delta \mu(\sigma)}
= \int_{\cal D} d{\bf r} \delta[\sigma - A_0({\bf r})]
\nonumber \\
\tilde g(\sigma) &=& -\frac{\delta {\cal F}}{\delta \tilde \mu(\sigma)}
= \int_{\cal D} d{\bf r} \{(\omega_0+f) \delta[\sigma - A_0({\bf r})]
\nonumber \\
&&-\ \gamma({\bf r},{\bf r}) \delta'[\sigma - A_0({\bf r})]
\label{9.15}
\end{eqnarray}
in which $\varepsilon({\bf r})$ is another microscale Gaussian fluctuation correction that may also be written terms of pair correlation functions \cite{W2012}. Note that due to continuity of $A$, $g(\sigma)$ is a large scale quantity, i.e., its own equilibrium average. Hence the level sets of a given initial condition $A({\bf r},t=0)$ are exactly preserved (though perhaps significantly contorted) in the equilibrium function $A_0({\bf r})$. On the other hand, due to strong (unbounded) fluctuations of $\omega$, a microscale correction to $\tilde g(\sigma)$ is evident.

The physically observable fields are the membrane gradients ${\bf B}$ and ${\bf v}$. Depending on the initial condition, their fluctuations, though bounded from point to point, could still be large compared to their mean values. This is physically quite different from the Euler equation where the second derivative has bounded fluctuations and the gradients are smooth. This has implications for the effects of dissipation which could be much stronger in this system, quelling micro-fluctuations and perhaps more rapidly bleeding energy out the large scale flow. The appearance (or not) of macroscale magnetic structure in the solar tachocline has significant implications for angular momentum transport between the two zones that it separates \cite{TDH2007}.

The example simulation result shown in Fig.\ \ref{fig:tachocline} is not intended as an equilibrium theory comparison---this will require future more careful study. However, it does verify that large scale magnetic field structures can survive for a long time even as the vorticity structure becomes much more diffuse. For this particular case the magnetic field magnitude is only weakly changed from its initial condition (not shown) while lack of vorticity conservation allows the zonal velocity magnitude to drop by nearly an order of magnitude.

\section{Conclusions}
\label{sec:conclude}

In this article we have discussed the application of statistical mechanics to the characterization of certain classes of large scale 2D steady state flows, following, for example, the free decay of an initial turbulent state (Fig.\ \ref{fig:turbmix}), highlighting the role of the competition between flow energy and microscale entropy production. The thermodynamic formalism makes sense only for systems whose dynamics is governed by a conserved Hamiltonian. When applied to fluid equations dynamics this limits consideration to idealized flows in which all dissipative terms are dropped. This, at minimum, limits the applicability to high Reynolds number flows with a large separation of scales between outer scale inertial, energy conserving dynamics, and small scaling mixing that eventually encounters viscous dissipation. With this separation, one may propose that the idealized models may provide reasonable predictions over an intermediate range of time scales that include a sufficient degree of intermediate scale equilibration.

This is especially interesting in two dimensional models, where one encounters an infinite number of conserved integrals of the motion (Casimirs) beyond the standard total energy and momentum. These strongly constrain the flow and in cases of interest lead to the phenomenon of an inverse cascade of energy to large scales, balanced by an ``enstrophy cascade'' to smaller scales, namely a fine-scale mixing of low energy eddies (Fig.\ \ref{fig:micromacro}). In a finite domain, the inverse cascade ``condenses'' into a system scale steady state structure. The goal of the thermodynamic treatment is to predict such structures based only on the values of the conserved integrals imposed by the initial flow---the only quantities ``visible'' to the statistical formalism. Given the very large number of such integrals, there are potentially many different large scale flow patterns that might be accessed, exemplified by long lived eddies such as Jupiter's Great Red Spot, zonal jet features, etc.

Following the classic construction of the statistical formalism (dynamics in phase space, ergodic hypothesis, Liouville theorem, invariant measures, choice of ensemble), the problem may be reduced to the analysis of a classical field theory (Table \ref{tab:fields}), with analogies to continuous spin Ising models (perhaps interacting with additional Gaussian degrees of freedom), and interacting elastic membrane models, depending on the exact problem and the fluid degree of freedom to which the Casimirs are applied. The fluid physics, however, drives these models into unusual regimes, e.g., of very high energy (negative temperatures) that are not normally encountered in more conventional versions of these models (Figs.\ \ref{fig:ptvortex} and \ref{fig:entropy}). In these regimes we have seen the statistical approach, in the form of a thermodynamic free energy variational principle, is indeed able to produce the desired macroscopic flows. The formalism additionally lends insight into the role of the various conserved integrals in controlling the geometry of these flows. Simple examples for the 2D Euler equation are shown in Fig.\ \ref{fig:2leveleg}.

Despite the mathematical elegance of the theory and its predictions, there remain numerous questions regarding the validity of the underlying assumptions, especially the ergodic hypothesis and the convergence to a true equilibrium state \cite{BV2012}. In comparison to conventional particle systems, there are many possible barriers to equilibration, including extra adiabatic invariants (Sec.\ \ref{sec:adiabatconserve}), metastable equilibria \cite{CC1996}, and very long-lived chaotic states (Sec.\ \ref{sec:eqissue}). Some of these are well understood, others deserve more careful study.

There are also systems for which the equilibrium theory apparently works too well! Thus, the inclusion of additional physical degrees of freedom intended to make the model more physically realistic, such as surface motions in the shallow water equations (Sec.\ \ref{sec:swwaveeddy}), in principle destabilizes negative temperatures states, leading to an ultraviolet catastrophe of surface waves despite the Casimir constraints. In fact, long-lived planetary eddies are much more in line with predictions of the much simpler Euler or quasigeostrophic equation \cite{BV2012}. Similar issues are seen in axisymmetric flows (Sec.\ \ref{sec:3daxisym}) where an ultraviolet catastrophe of poloidal vorticity predicts only rather trivial large scale toroidal flows. In both cases the catastrophic coupling of the new small scale fluctuations to existing large scale structures is likely very weak, and high frequency wave or poloidal vorticity generation might better be thought of as an additional weak dissipation mechanism that can also be ignored over time scales of interest. The resulting quasi-hydrostatic limit of the shallow water equations provides one possible route to formally maintaining negative temperature states while still treating the surface height in a consistent manner (Sec.\ \ref{sec:quasihydrosweq}).

The previous discussion motivates a number of future investigations into a more careful treatment of additional dynamical time- and length-scale separations that could either hinder or aid statistical equilibrium approximations, and how to properly define the effective conserved integrals entering a new idealized flow model, e.g., through an appropriate spatial filter.

In addition, it is clear that very long-lived eddies, such as Jupiter's Red Spot, require some sort of driving force to survive. The weak driving--weak dissipation limit could perhaps be formulated through convergence to a near-equilibrium state in which the conserved integrals come into detailed balance, e.g., through some kind of Onsager nonequilibrium linear response theory applied to the fluid Hamiltonian. On the other hand, it is known that weak stochastic forcing can occasionally lead to rare, sudden, catastrophic changes to the state \cite{LB2014,BLZ2014} so some care must be taken in finding the correct regime in which to formulate the problem.


\begin{thebibliography}{}
\bibitem{foot:ablim} This can be made rigorous by letting both the intermediate scale $b$ and the grid scale $a \to 0$, but with $b/a \to \infty$ \cite{MWC1992}. In this way the local integral of $\omega - \omega_0$ against the smooth function $\psi_0$ in (\ref{4.2}) scales as $a/b \to 0$. Thus, the Casimir constraints ensure that although $\omega$ is discontinuous from grid point to grid point on the microscale $a$, its local fluctuations are bounded by the support of $g(\sigma)$. It follows also that the velocity ${\bf v}({\bf r})$ is continuous and the stream function $\psi$ is continuously differentiable.
\bibitem{foot:rossby} One may write $c = \sqrt{g_\mathrm{eff} H_\mathrm{eff}}$ where $H_\mathrm{eff}$ is the effective fluid layer depth and $g_\mathrm{eff}$ the effective acceleration due to gravity. For single layer shallow water theory, $g_\mathrm{eff} = g$ and $H_\mathrm{eff} = H$ are the ``bare'' physical values. For internal waves in a density stratified medium $g_\mathrm{eff}/g \propto \delta \rho/\rho$ is reduced by the density constrast between layers, and $H_\mathrm{eff}$ is the scale height, namely the effective height of the water column actually taking part in the motion (e.g., thermocline depth). In the Earth's oceans, $\delta \rho/\rho \sim 10^{-2}$ and the Kelvin wave speed $c \sim 2$ m/s is therefore $O(100)$ times smaller than the ``bare'' shallow water wave speed.
\end{thebibliography}
\end{document}